\documentclass[sigconf, nonacm]{acmart}

\newcommand\vldbdoi{XX.XX/XXX.XX}
\newcommand\vldbpages{XXX-XXX}
\newcommand\vldbvolume{14}
\newcommand\vldbissue{1}
\newcommand\vldbyear{2020}
\newcommand\vldbauthors{\authors}
\newcommand\vldbtitle{\shorttitle} 
\newcommand\vldbavailabilityurl{URL_TO_YOUR_ARTIFACTS}
\newcommand\vldbpagestyle{plain}

\usepackage{amsmath,amsfonts}
\usepackage{algorithmic}
\usepackage{graphicx}
\usepackage{textcomp}
\usepackage{hyperref}
\usepackage{xcolor}
\usepackage{url}
\usepackage{float}
\usepackage{multirow}
\usepackage[most]{tcolorbox}
\tcbuselibrary{breakable} % <-- ADD THIS LINE
\usepackage{listings}
\usepackage{array}
\usepackage{booktabs}
\usepackage{enumitem}
\usepackage{fontawesome5}
\usepackage{tikz}
\usetikzlibrary{positioning, fit, shapes.geometric, shapes.symbols, arrows.meta, backgrounds, shadows}

\definecolor{lstpurple}{rgb}{0.5,0,0.5}
\definecolor{lstblue}{RGB}{0,0,255}
\definecolor{lstgreen}{RGB}{0,128,0}
\definecolor{lstreddark}{RGB}{180,0,0}

\lstdefinestyle{SQL}{
  language=SQL,
  basicstyle=\small\ttfamily,
  keywordstyle=\color{blue}\bfseries,
  commentstyle=\color{gray},
  stringstyle=\color{orange},
  numbers=left,
  numberstyle=\tiny,
  stepnumber=1,
  numbersep=5pt,
  backgroundcolor=\color{white},
  frame=lines,
  tabsize=4,
  showspaces=false,
  showstringspaces=false
}

\tcbuselibrary{listings,skins}
\tcbset{
  SQLbox/.style={
    enhanced,
    colback=white,
    colframe=blue!20!black,
    sharp corners,
    frame hidden,
    interior hidden,
    boxsep=0pt,
    top=0pt,
    bottom=0pt,
  }
}

\newtheorem{Example}{Example}

\lstdefinestyle{psqlcolor}
{
columns=fullflexible,
tabsize=2,
basicstyle=\footnotesize\upshape\ttfamily,
language=SQL,
extendedchars=false,
keywordstyle=\bfseries\color{lstpurple},
deletekeywords={count,min,max,avg,sum},
keywordstyle=[2]\color{lstblue},
stringstyle=\color{lstreddark},
commentstyle=\color{lstgreen},
mathescape=true,
escapechar=@,
sensitive=true,
numbers=left,
breaklines=true,          % <--- ADDS AUTO WORD WRAP
breakatwhitespace=true,   % <--- ONLY BREAKS AT SPACES
postbreak=\mbox{\textcolor{gray}{$\hookrightarrow$}\space} % <--- ADDS WRAP ARROW
}

\newcommand{\attr}{A}
\newcommand{\rel}{R}
\newcommand{\cons}{C}
\newcommand{\schema}{S}
\newcommand{\model}{\texttt{M}}
\newcommand{\workload}{\mathcal{W}}

\newcommand{\mypipeline}{\textsc{BaseSQL}}
\begin{document}

%% Title
\title{The Case for Text-to-SQL Friendly Logical Database Design}

%% Authors
\author{Shi Heng Zhang}
\email{andy_zhang@sfu.ca}
\affiliation{%
  \institution{Simon Fraser University}
  \country{Canada}
}

\author{Zhengjie Miao}
\email{zhengjie@sfu.ca}
\affiliation{%
  \institution{Simon Fraser University}
  \country{Canada}
}

\author{Jiannan Wang}
\email{jnwang@tsinghua.edu.cn}
\affiliation{%
  \institution{Tsinghua University}
  \country{China}
}

%% Abstract
\begin{abstract}

Logical database design has traditionally optimized database schemas, including tables, columns, keys, constraints, and views, for correctness, integrity, and human-written application queries. LLM-based Text-to-SQL changes the consumer: the schema is now often read as text by a language model, so design choices that preserve database semantics can still change SQL-generation accuracy. We argue that this creates a new design objective alongside the classical ones --- \emph{LLM-friendly logical database design}, the property that a schema is easy for a language model to map from natural language to correct SQL --- and treat it as the optimization target of this paper. We instantiate this objective with three semantics-preserving schema transformations that re-purpose classical schema-design ideas: schema abstraction (\texttt{+A:} logical views that materialize recurring join paths), schema partitioning (\texttt{+P:} workload-aware logical partitions that prune irrelevant context), and schema renaming (\texttt{+R:} descriptive identifiers that improve downstream column linking and predicate construction). The three operators compose, and each preserves the underlying database semantics. When historical question--SQL pairs are available, they guide both partitioning and abstraction; in zero-shot settings, renaming applies directly, and abstraction falls back to an ad-hoc per-question variant. We evaluate the resulting schemas on BIRD-Union and Spider-Union across multiple Text-to-SQL pipelines and language model backbones, with gains of up to 4.2\% in execution accuracy. The best transformation varies modestly across pipelines and models, with the full \texttt{+A+P+R} consistently improving; multiple operator combinations are competitive on each pipeline. These results show that LLM-friendly logical design is a practical and underexplored database-side optimization target, complementary to existing Text-to-SQL pipelines.
\end{abstract}

\pagestyle{plain}
\maketitle

\iffalse % <--- ADD THIS TO HIDE THE BLOCKS FOR ARXIV
\pagestyle{\vldbpagestyle}
\begingroup\small\noindent\raggedright\textbf{PVLDB Reference Format:}\\
\vldbauthors. \vldbtitle. PVLDB, \vldbvolume(\vldbissue): \vldbpages, \vldbyear.\\
\href{https://doi.org/\vldbdoi}{doi:\vldbdoi}
\endgroup
\begingroup
\renewcommand\thefootnote{}\footnote{\noindent
This work is licensed under the Creative Commons BY-NC-ND 4.0 International License. Visit \url{https://creativecommons.org/licenses/by-nc-nd/4.0/} to view a copy of this license. For any use beyond those covered by this license, obtain permission by emailing \href{mailto:info@vldb.org}{info@vldb.org}. Copyright is held by the owner/author(s). Publication rights licensed to the VLDB Endowment. \\
\raggedright Proceedings of the VLDB Endowment, Vol. \vldbvolume, No. \vldbissue\ %
ISSN 2150-8097. \\
\href{https://doi.org/\vldbdoi}{doi:\vldbdoi} \\
}\addtocounter{footnote}{-1}\endgroup

\ifdefempty{\vldbavailabilityurl}{}{
\vspace{.3cm}
\begingroup\small\noindent\raggedright\textbf{PVLDB Artifact Availability:}\\
The source code, data, and/or other artifacts have been made available at \url{https://anonymous.4open.science/r/Logical-Database-Design-CC0C}.
\endgroup
}
\fi % <--- ADD THIS TO CLOSE THE HIDDEN SECTION

%\vspace{-2mm} % Adjust from -1mm to -5mm until it pops over
\section{Introduction}

Logical database design depicts how real-world entities and relationships are organized into a relational database~\cite{garcia-molina-dbsystems}. Traditionally, this process focuses heavily on system efficiency and optimization, prioritizing normal forms, data integrity, and eliminating redundancy over schema semantic clarity. While system efficiency is vital, the resulting schema is still the primary interface for consumption by developers. A well-designed schema should provide clear and domain-aligned abstractions, making it easier to reason about the underlying data and produce the correct SQL queries~\cite{snail, furst2025evaluating}.

%nl2sql emerged, so the consumer changed, new options for logical design. currently... only a few works explained the impacts of schema on nl2sql, but not frame it as a design problem -- Text-to-SQL has advanced rapidly in recent years, driven largely by large language models (LLMs) and increasingly sophisticated pipelines, including the modules of schema linking, iterative self-correction, candidate generation, reranking, and retrieval over large schemas. Recent systems such as CHESS~\cite{talaei2025chess}, OpenSearch-SQL~\cite{xie2025opensearchsql}, RSL-SQL~\cite{cao2024rsl}, RASL~\cite{eben2025rasl}, and TS-SQL~\cite{xu2025tssql} have substantially improved execution accuracy on challenging benchmarks such as BIRD~\cite{li2023bird}, alongside emerging frameworks that introduce advanced candidate generation and decomposed pipelines~\cite{alphasql2025, reforce2025, deepeyesql2026}. Despite this progress, however, most existing work treats the database schema as fixed input and focuses on adapting the model, prompt, or decoding process to that schema. As a matter of fact, recent studies on schema robustness (e.g., EvoSchema~\cite{evoschema2025}) demonstrate that current LLMs remain highly fragile to underlying logical design choices and schema evolution, highlighting the urgent need for schema-level optimization.

In recent years, Text-to-SQL has emerged as a powerful tool to aid the query-writing process, allowing users to interact with databases using natural language questions. Driven by the rapid advancement of large language models (LLMs), this shift has fundamentally changed the consumption of the database schema: instead of being consumed exclusively by developers, the schema is now read and interpreted as textual context by the LLM. Modern Text-to-SQL systems ~\cite{talaei2025chess, xie2025opensearchsql, cao2024rsl, eben2025rasl, xu2025tssql} have substantially improved execution accuracy on challenging benchmarks like BIRD~\cite{li2023bird}. Despite this progress, most existing work treats the database schema as a fixed input, focusing exclusively on adapting the model, prompt, or decoding process to that schema. Recent empirical studies on schema robustness (e.g., EvoSchema~\cite{evoschema2025}) demonstrate that LLMs remain highly fragile to underlying logical design choices and schema evolution, highlighting the urgent need for schema-level optimization.

In this paper, we argue that this shift introduces a new optimization objective for logical database design: \emph{Text-to-SQL-friendly logical design}---the property that a schema is easy for a language model to map from natural language to correct SQL. This objective operates independently of classical design goals, such that schemas that model the same underlying data would differ drastically in their Text-to-SQL-friendliness. For example, highly normalized schemas often require multi-table joins and bridge tables even for relatively simple questions~\cite{furst2025evaluating}; weak or ambiguous identifiers can hinder lexical alignment between the question and the schema; and large enterprise schemas introduce substantial irrelevant context that confuses schema linking and retrieval. Ultimately, the same database schema varies in Text-to-SQL-friendliness depending on how its logical structure is presented to the model.

%This observation motivates a complementary, \emph{database-centric} perspective on Text-to-SQL. Rather than only asking how to improve the model or system for a given schema, we ask whether the schema representation itself can be transformed to better support Text-to-SQL generation. In particular, can we construct a semantically faithful but more \emph{Text-to-SQL-friendly} logical design that reduces unnecessary reasoning burdens, improves lexical and semantic alignment, and suppresses irrelevant schema noise?

This observation motivates a complementary, \emph{database-centric} perspective on Text-to-SQL. Rather than only asking how to improve the model for a given schema, we ask whether the schema itself can be transformed to better support the model. Specifically, given a database schema and a historical Text-to-SQL workload (optional), our goal is to construct a transformed schema that increases downstream Text-to-SQL accuracy.

\begin{figure*}[t]
    \centering
    \vspace{-5pt} % Pulls the diagram up closer to the paragraph above it
    
    % trim = left bottom right top
    % You MUST include 'clip', otherwise the white space is still there, just invisible!
    \includegraphics[page=5, trim=0cm 3.3cm 0cm 0.7cm, clip, width=0.85\textwidth]{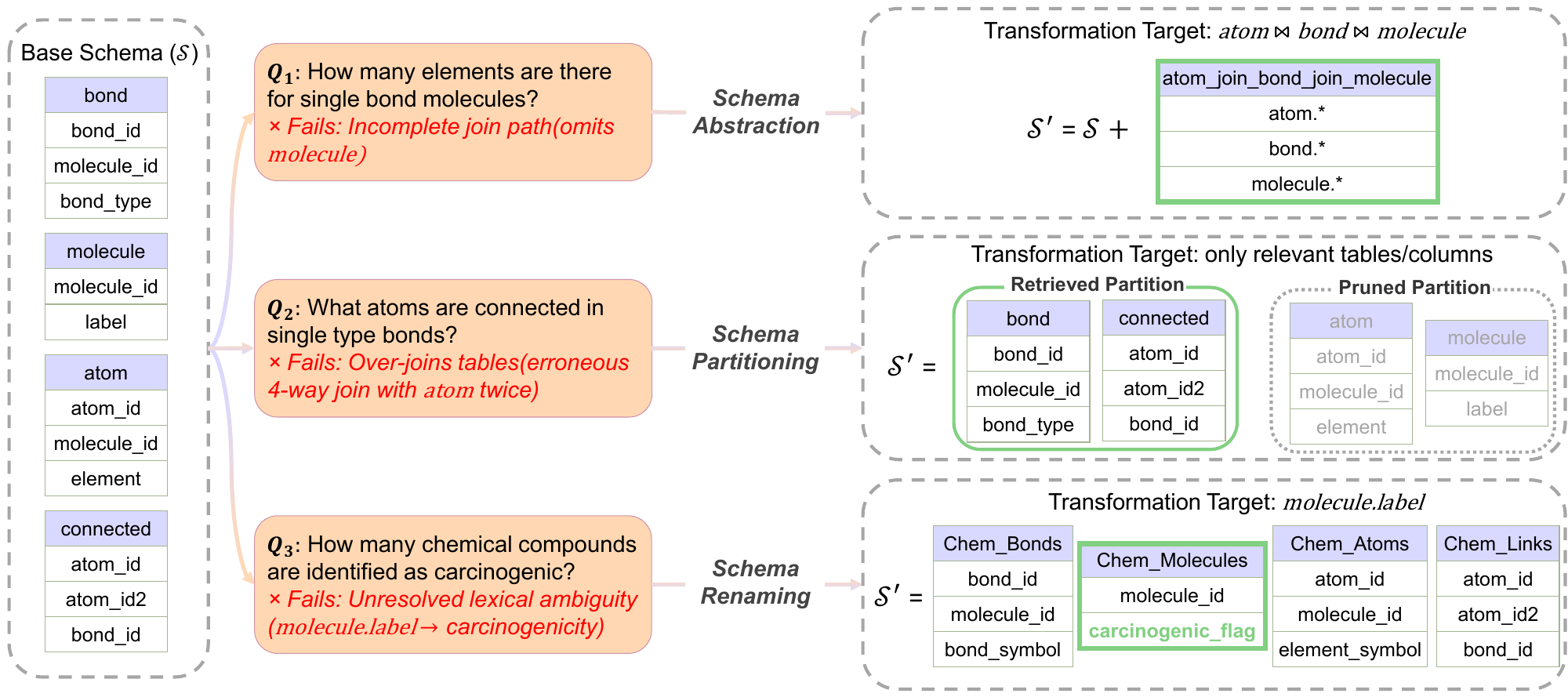}
    
    \vspace{-5pt} % Snugs the caption slightly closer to the diagram
    \caption{Running example illustrating how schema transformations mitigate complementary Text-to-SQL failure modes.}
    \label{fig:running_example}
    
    \vspace{-12pt} % Aggressively pulls the 5.5 heading up closer to the caption
\end{figure*}

%A major source of difficulty arises from \textbf{schema-induced reasoning complexity}. In many cases, Text-to-SQL systems fail before higher-level reasoning even begins: they select the wrong tables, infer incomplete join paths, or introduce unnecessary joins. These errors are especially common in normalized schemas, where the model must recover latent relational structure before it can express the intended query. More generally, weak identifiers and large schema contexts can further hinder schema linking and retrieval. From a database perspective, these challenges are closely related to a broader design question: how should schemas and workload-relevant abstractions be organized so that recurring access patterns become easier to express? While such concerns are central in classical database design and workload-driven optimization, they have received much less attention in Text-to-SQL.

To achieve this, we analyze the underlying process by which a language model reads and uses a database schema. To successfully generate a query, the model needs to understand three core dimensions of the schema: its relational structure (how tables connect), its relevance (which tables matter), and its semantics (what the tables and columns mean). To help the model navigate these three dimensions, we propose three corresponding transformations: \textbf{Schema Abstraction (+A)} simplifies the relational structure by exposing recurring joined paths as logical views, \textbf{Schema Partitioning (+P)} organizes the schema into localized partitions to keep only the relevant schema and historical workload, and \textbf{Schema Renaming (+R)} reduces ambiguity by substituting table and column names with more descriptive identifiers. By applying these three operators, we adapt classic database optimization techniques to help language models achieve higher query execution accuracy.

\vspace{-2mm} % <--- Adjust this number (-3mm, -5mm) until the gap looks perfect
\begin{Example}[Running example]
Consider a molecular database, adapted from the BIRD benchmark~\cite{li2023bird}, with relations such as
\texttt{molecule}, \texttt{atom}, \texttt{bond}, and \texttt{connected}, as shown in Figure~\ref{fig:running_example}. Although this schema is reasonable from a normalized relational-design perspective, it can be difficult for a Text-to-SQL model to use directly.

\noindent To illustrate \textbf{Schema Abstraction (+A)}, consider question $Q_1$ in Figure~\ref{fig:running_example}: \textit{``How many elements are there for single bond molecules?''} The intended query requires the model to join \texttt{atom}, \texttt{bond}, and \texttt{molecule}. A baseline model may fail to produce this full path, generating an incomplete query over only \texttt{atom} and \texttt{bond}, e.g.,
\[
\texttt{atom} \Join \texttt{bond}
\]
thereby omitting the required context by the question. By applying +A, the framework instead exposes a semantic view,
\[\texttt{atom\_join\_bond\_join\_molecule},\]
so that the model can express the query as a simple selection and aggregation over a single higher-level relation.

\noindent To demonstrate \textbf{Schema Partitioning (+P)}, consider question $Q_2$: \textit{``What atoms are connected in single type bonds?''} The correct query only needs the relationship between \texttt{bond} and \texttt{connected}, projecting the \texttt{connected} atom identifiers:
\begin{tcolorbox}[SQLbox, before skip=4pt, after skip=4pt]
\begin{lstlisting}[style=psqlcolor,mathescape,deletendkeywords={YEAR, DATE}]
SELECT connected.atom_id, connected.atom_id2 
FROM bond JOIN connected ON ...
\end{lstlisting}
\end{tcolorbox}
In contrast, the baseline model may over-join schema elements and introducing an incorrect four-way join with \texttt{atom} twice, while also not correctly projecting the \texttt{connected} atom identifiers. By applying +P, we restrict the context strictly to the relevant partition,
\[
\{\texttt{bond}, \texttt{connected}\},
\]
and retrieve demonstrations that reference this exact partition, guiding the model toward the correct join and projection patterns.

\noindent Finally, to show the impact of \textbf{Schema Renaming (+R)}, consider question $Q_3$, which states  ``How many chemical compounds are identified as carcinogenic?''. This information is in an ambiguous column \texttt{molecule.label}, making it difficult for the model to infer that this column is needed in a carcinogenicity predicate. By applying +R, we expose a clearer, semantically enriched identifier, e.g.,
%\vspace{-5mm} % <--- Adjust this number (-3mm, -5mm) until the gap looks perfect
\[
\texttt{molecule.label} \rightarrow \texttt{Chem\_Molecules.carcinogenic\_flag},
\]
%\vspace{-4mm} % <--- Adjust this number (-3mm, -5mm) until the gap looks perfect
consequently, establishing a direct lexical alignment between the natural-language question and the schema identifier.
\end{Example}
\vspace{-2mm} % <--- Adjust this number (-3mm, -5mm) until the gap looks perfect
The above example illustrates our core intuition: many Text-to-SQL failures are consequences of how the database schema is exposed to the model. This motivates \emph{Text-to-SQL-friendly logical database design}: an additional design objective that asks how the model-facing schema should be organized so that an LLM can more reliably produce the correct SQL from the natural-language questions. This objective complements classical logical-design goals such as enforcing normal forms, ensuring data integrity, and application support, but optimizes for a new consumer of the schema: the language model. Ultimately, it seeks to make schema semantics clearer, reduce irrelevant context, and expose recurring access patterns as higher-level abstractions.

To show that this objective is actionable, we instantiate it as \textbf{Logical Database Design for Text-to-SQL}: a database-side framework that transforms the model-facing schema through three semantics-preserving operators: schema abstraction(\texttt{+A}), schema partitioning(\texttt{+P}), and schema renaming(\texttt{+R}). These operators revisit classical schema-design ideas under a new optimization target: improving Text-to-SQL generation accuracy. The framework supports both zero-shot and workload-aware settings and can be integrated with existing Text-to-SQL pipelines as a database-side enhancement, without changing model weights or model architecture.

% Motivated by this perspective, we introduce \textbf{Logical Database Design for Text-to-SQL}, a database-side optimization framework that transforms the schema representation exposed to the model.  The key idea is to shift part of the reasoning burden from the model's inference process into the database design itself. Instead of repeatedly asking the model to infer useful schema structure from raw base tables, ambiguous identifiers, and large schema contexts, we expose a transformed logical representation that is easier to align with natural-language questions while remaining faithful to the original database semantics. Our framework supports both zero-shot and workload-aware settings and can be integrated with existing Text-to-SQL pipelines as a database-side enhancement.

% Experiments on BIRD-Union and Spider-Union show that logical schema transformation improves Text-to-SQL performance in both zero-shot and history-aware settings. The gains are not uniform across all pipelines, but selected transformations consistently improve strong baselines, with the largest improvements arising when schema-related bottlenecks such as join complexity, weak identifier semantics, or excessive schema scope are prominent. These results suggest that logical database design is an important and underexplored dimension for Text-to-SQL.

Our main contributions are as follows:

\begin{itemize}[leftmargin=*]
    % \item \textbf{A database-centric perspective on Text-to-SQL.}
    % We argue that logical schema design itself is an important optimization dimension for Text-to-SQL, complementing existing model-side and query-side approaches.
    \item \textbf{A new optimization objective for logical schema design.}
    We identify and study \emph{Text-to-SQL-friendly logical design}: the property that a model-facing schema makes it easier for a language model to map natural-language questions to correct SQL. We argue that this should be treated as an additional logical-design objective alongside classical goals.

    \item \textbf{Schema-design ideas for Text-to-SQL.}
    We instantiate this objective through three semantics-preserving transformations that repurpose classical schema-design ideas for the new consumer: schema abstraction, which exposes recurring joins as logical views; schema partitioning, which organizes the schema into workload-aware partitions; and schema renaming, which improves identifier clarity and lexical alignment.

    \item \textbf{An operational framework with zero-shot and history-aware modes.}
    We develop a database-side framework that applies these transformations without changing the underlying database or the Text-to-SQL model. 
    It supports workload-aware partitioning and few-shot demonstration retrieval when history is available, and schema-only renaming and ad-hoc abstraction when history is absent. 

    \item \textbf{Comprehensive empirical evaluation.}
     Experiments on BIRD-Union and Spider-Union show that Text-to-SQL-friendly logical schema design consistently improves strong Text-to-SQL baselines across multiple LLM backbones and pipeline architectures, and reveals where the gains concentrate.
    
    % \item \textbf{A framework for Text-to-SQL-friendly logical design.}
    % We introduce a database-side optimization framework that transforms the schema representation exposed to the model while preserving the semantics of the underlying database.
    
    % \item \textbf{Three classes of schema transformations.}
    % We develop a modular transformation space consisting of metadata enrichment and schema renaming, schema partitioning with workload-aware organization, and schema abstraction.

    % \item \textbf{Support for both zero-shot and history-aware settings.}
    % Our framework operates without historical workloads, but can also exploit workload history to improve schema partitioning and retrieval when such information is available.

    % \item \textbf{Comprehensive empirical evaluation.}
    % Experiments on BIRD-Union and Spider-Union show that logical schema transformation improves strong Text-to-SQL baselines across both zero-shot and history-aware settings, while also revealing that the best transformation is pipeline-dependent.
\end{itemize}
\section{RELATED WORK}

\noindent\textbf{Traditional Text-to-SQL Approaches.}
While early encoder decoder models established the task, today’s state-of-the-art relies on LLMs, sophisticated prompting, and self-repair. Prompt-centric approaches like DIN-SQL~\cite{pourreza2023dinsql} and DAIL-SQL~\cite{gao2024dailsql} improve performance through modular decomposition and execution-guided reasoning. To tackle enterprise-level databases, systems such as RSL-SQL~\cite{cao2024rsl}, OpenSearch-SQL~\cite{xie2025opensearchsql}, CHESS~\cite{talaei2025chess}, and RASL~\cite{eben2025rasl} explicitly ground the LLM using schema linking and retrieval augmentation. Concurrently, the field has advanced through complex reasoning architectures, including Monte Carlo Tree Search~\cite{alphasql2025}, test-driven refinement~\cite{xu2025tssql}, and multi-agent pipelines leveraging corrective self-consistency~\cite{deepeyesql2026, wang2025mac, CSCSQL, reforce2025}. Despite these tremendous advancements, existing systems treat the database schema as a static input, which motivates our database-centric approach of dynamically optimizing the logical schema representation.

\smallskip
\noindent\textbf{Importance of Schema for Text-to-SQL.}
A smaller but growing body of work recognizes that modifying the \emph{schema} heavily alters downstream Text-to-SQL performance. A recent empirical study, SNAILS~\cite{snail}, demonstrates that semantically richer column and table identifiers can boost LLM accuracy by 5–10\%. Similarly, EvoSchema~\cite{evoschema2025} highlights that current LLMs are very susceptible to schema evolution and logical design shifts. Other efforts to bridge this semantic gap include expanding schema aliases and pruning~\cite{zhao2022schemaexpansion}, attaching rich attribute annotations to DDL prompts~\cite{bmc22022sqlcreatecontext}, and generating synthetic foreign keys~\cite{li2025omnisql}. While these works validate the importance of schema semantics, none of them suggest pre-computing joins as views to systematically apply dynamic logical transformations to augment the schema search space.

\smallskip
\noindent\textbf{View Recommendation and Physical Design.}
View selection is a classic problem in physical database design. Foundational work~\cite{harinarayan1996implementing} and systems like AutoAdmin~\cite{chaudhuri1997autoadmin} frame it as a cost-benefit optimization, trading maintenance overhead for query savings under storage constraints~\cite{chaudhuri2007selftuning}.  Later systems such as DynaMat~\cite{kotidis1999dynamat}, Microsoft’s automated design tools~\cite{agrawal2000automated}, and the DB2 Design Advisor~\cite{zilio2004db2} extended these ideas using workload-aware what-if costing and integer programming. While these traditional systems automate design to optimize \textit{query execution}, we adopt this same philosophy to optimize \textit{query generation} for the language model.

\smallskip
\noindent\textbf{Workload-Aware Schema Adaptation.}
Building on the foundations of index and view selection, the database community has long utilized historical query workloads to drive structural schema optimizations. Examples include automatic vertical partitioning for OLTP workloads (e.g., AutoPart~\cite{agrawal2004autopart}) and graph-based data placement to minimize distributed joins (e.g., Schism~\cite{curino2010schism}). Moreover, the vision of autonomous database systems~\cite{pavlo2017selfdriving} encourages monitoring query access patterns to dynamically restructure the underlying storage and logical layout. Traditionally, these systems utilize workload analysis exclusively to optimize query execution performance.
\section{Problem Formulation}
\label{sec:problem_formulation}

\subsection{Logical Database Design Meets Text-to-SQL}
Logical database design creates a relational schema --- tables, columns, data types, and integrity constraints --- guided by entity-relationship modeling and normalization theories to reduce redundancy and preserve integrity. In Text-to-SQL, this schema is also part of the input that the generative model conditions on: the same design choices that impact storage and integrity also dictate how easily a natural-language question can be expressed as SQL. Yet the effect of logical design on Text-to-SQL accuracy remains under-explored. For human developers, query writing becomes considerably easier when the query requires fewer joins under the given schema, uses self-descriptive identifiers, and demands little external knowledge; we hypothesize that the same factors translate to Text-to-SQL systems, which increasingly mimic human query-writing patterns.

\begin{figure*}[t]
\centering
\resizebox{0.95\linewidth}{!}{%
\begin{tikzpicture}[
    >=Stealth,
    node distance=0.4cm and 1.2cm,
    font=\sffamily,
    % Modern UI Card Styles
    db/.style={rectangle, rounded corners=6pt, draw=gray!30, thick, fill=white, drop shadow={opacity=0.08, shadow xshift=1pt, shadow yshift=-1pt}, minimum width=2.6cm, minimum height=1.1cm, align=center},
    optschema/.style={rectangle, rounded corners=6pt, draw=green!40!black, thick, fill=green!2, drop shadow={opacity=0.1, shadow xshift=1pt, shadow yshift=-1pt}, minimum width=2.6cm, minimum height=1.1cm, align=center},
    mod/.style={rectangle, rounded corners=8pt, draw=blue!30, thick, fill=blue!2, drop shadow={opacity=0.1, shadow xshift=1pt, shadow yshift=-1pt}, minimum width=4.6cm, minimum height=0.9cm, align=center},
    llm/.style={rectangle, rounded corners=6pt, draw=purple!30, thick, fill=purple!2, drop shadow={opacity=0.1, shadow xshift=1pt, shadow yshift=-1pt}, minimum width=3.2cm, minimum height=0.9cm, align=center},
    io/.style={rectangle, rounded corners=6pt, draw=teal!30, thick, fill=white, drop shadow={opacity=0.08, shadow xshift=1pt, shadow yshift=-1pt}, minimum width=2.2cm, minimum height=0.8cm, align=center},
    % Soft Grouping Boxes
    phasebox/.style={rectangle, rounded corners=10pt, draw=gray!20, dashed, thick, fill=gray!3, inner sep=10pt},
    enginebox/.style={rectangle, rounded corners=10pt, draw=blue!15, thick, fill=white, inner sep=8pt},
    grouptitle/.style={font=\bfseries\sffamily\small, text=gray!60!black}
]

% ------------------------------------------
% PERFECT ALIGNMENT ANCHORS
% ------------------------------------------
% Strictly locks BOTH boxes to an identical, compact height
\coordinate (box_top) at (0, 2.0);
\coordinate (box_bot) at (0, -2.0);

% ==========================================
% PHASE 1: OFFLINE SCHEMA OPTIMIZATION
% ==========================================

% The Modules (Reordered to +A, +P, +R)
\node[mod] (cluster) {\textbf{\textcolor{green!60!black}\ +P: Schema Partitioning} \\[-2pt] \scriptsize Prunes Contextual Noise};
\node[mod, above=0.2cm of cluster] (view) {\textbf{\textcolor{green!60!black}\ +A: Schema Abstraction} \\[-2pt] \scriptsize Bypasses Complex Joins};
\node[mod, below=0.2cm of cluster] (rename) {\textbf{\textcolor{green!60!black}\ +R: Schema Renaming} \\[-2pt] \scriptsize Resolves Lexical Ambiguity};

% The Engine Box
\begin{scope}[on background layer]
    \node[enginebox, fit=(rename) (view) (cluster)] (engine) {};
\end{scope}

% The Title (Placed directly ON the blue border like a fieldset legend)
\node[grouptitle, text=blue!80!black, fill=white, inner sep=4pt] at (engine.north) {Modular Transformation Pool $g(\mathcal{\schema}, \mathcal{H})$};

% The Inputs (Stacked vertically, symmetrically aligned to the NEW center module)
\node[db, left=2.0cm of cluster, yshift=0.7cm] (baseschema) {\faDatabase\\[1pt] Base Schema $\mathcal{\schema}$};
\node[db, left=2.0cm of cluster, yshift=-0.7cm] (history) {\faHistory\\[1pt] Workload $\mathcal{H}$};

% The Output (Aligned to the NEW center module)
\node[optschema, right=1.4cm of cluster] (optschema) {\faDatabase\\[1pt] Optimized Schema $\mathcal{\schema}'$};

% Smooth Data Streams - Merging Inputs into a Center Point shifted left (closer to boxes)
\coordinate (input_joint) at ([xshift=0.6cm]baseschema.east |- cluster.west);

% Merging Inputs with strict bezier curves to prevent "snakey" bulging
\draw[thick, draw=gray!40] (baseschema.east) .. controls +(0.3, 0) and +(-0.3, 0) .. (input_joint);
\draw[thick, draw=gray!40] (history.east) .. controls +(0.3, 0) and +(-0.3, 0) .. (input_joint);

% Fanning out to the Modules with strict bezier curves
\draw[->, thick, draw=gray!40] (input_joint) .. controls +(0.6, 0) and +(-0.6, 0) .. (view.west);
\draw[->, thick, draw=gray!40] (input_joint) -- (cluster.west); % Direct straight line to the center module
\draw[->, thick, draw=gray!40] (input_joint) .. controls +(0.6, 0) and +(-0.6, 0) .. (rename.west);

% Smooth Data Streams - Flowing out to Optimized Schema
\draw[->, thick, draw=blue!40] (view.east) to[out=0, in=180] (optschema.west);
\draw[->, thick, draw=blue!40] (cluster.east) to[out=0, in=180] (optschema.west);
\draw[->, thick, draw=blue!40] (rename.east) to[out=0, in=180] (optschema.west);

% Phase 1 Background Box (Locked to Anchors)
\begin{scope}[on background layer]
    \node[phasebox, fit=(baseschema) (history) (engine) (optschema) (engine |- box_top) (engine |- box_bot)] (offlinebox) {};
\end{scope}
\node[grouptitle, above=2pt of offlinebox.north] {Offline Schema Transformation};

% ==========================================
% PHASE 2: ONLINE INFERENCE
% ==========================================

% Online Pipeline components
\node[llm, right=1.6cm of optschema] (llm) {\faRobot\\[1pt] Text-to-SQL Pipeline $\mathcal{M}$\\ \scriptsize (e.g., \mypipeline{}, DIN-SQL)};
% Gap set to 0.35cm to evenly distribute the stack within the ±2.0 height
    \node[io, above=0.35cm of llm] (question) {\faUser\\[1pt] User Question $q$};
\node[io, below=0.35cm of llm] (sql) {\faCode\\[1pt] Generated SQL $y$};

% Arrows (Vertical flow + Horizontal Schema Injection)
\draw[->, thick, draw=gray!50] (question.south) -- (llm.north);
\draw[->, thick, draw=green!60!black] (optschema.east) -- node[above, font=\scriptsize] {Plug \& Play} (llm.west);
\draw[->, thick, draw=purple!50] (llm.south) -- (sql.north);

% Phase 2 Background Box (Locked to Anchors)
\begin{scope}[on background layer]
    \node[phasebox, fit=(question) (llm) (sql) (llm |- box_top) (llm |- box_bot)] (onlinebox) {};
\end{scope}
\node[grouptitle, above=2pt of onlinebox.north] {Online Prompt Assembly};
\end{tikzpicture}%
}
\vspace{-2mm}
\caption{The proposed database-centric Text-to-SQL framework. The transformation function $g$ utilizes offline schema transformation and online prompt assembly to construct an optimized representation $\mathcal{S}'$ for any downstream pipeline $\mathcal{M}$.}
\label{fig:pipeline}
\vspace{-4mm}
\end{figure*}
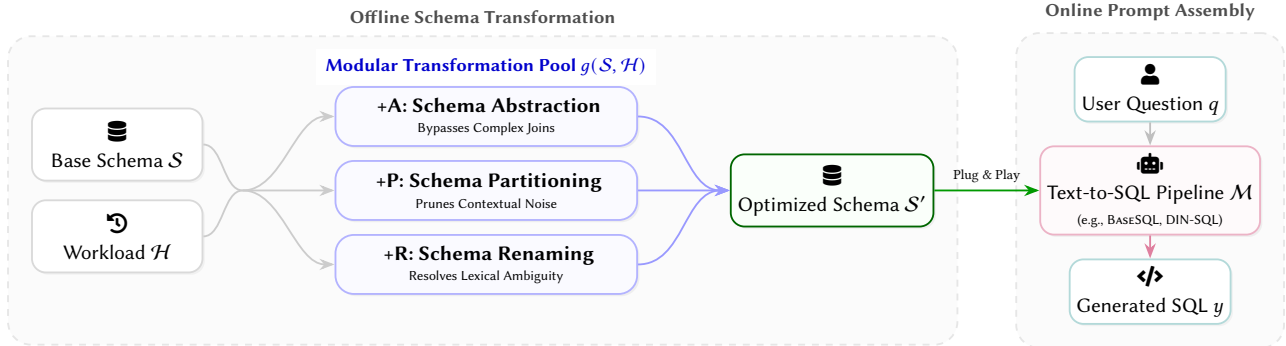

\subsection{Text-to-SQL-Friendly Database Design as an Optimization Problem}

\noindent\textbf{Database and Schema.} Let $\mathcal{D}$ be a relational database with schema \[
\mathcal{\schema} = (\mathcal{\rel}, \mathcal{\attr}, \mathcal{\cons}),
\] where $\mathcal{\rel} = \{\rel_1,\dots,\rel_n\}$ is the set of relations, $\mathcal{\attr}$ is the set of attributes, and $\mathcal{\cons}$ represents the integrity constraints, including primary-key and foreign-key constraints.

\noindent\textbf{The Task.} Let $\model$ be a text-to-SQL system. Given a natural-language question $q$ and a schema $\mathcal{\schema}$, the system generates a SQL query \[y = \model(q, \mathcal{\schema}).\]

Let $\workload$ denote a target workload of user questions. Optionally, let $\mathcal{H}$ denote historical workload associated with the same database, such as prior questions and query logs. In the zero-shot setting, $\mathcal{H}$ may be empty. We compute $g$ from $(\mathcal{\schema}, \mathcal{H})$ at offline time and evaluate accuracy on a held-out workload $\workload$ disjoint from $\mathcal{H}$. In the few-shot regime, $\model$ additionally receives top-$k$ demonstrations $D_q \subseteq \mathcal{H}$ selected by a retrieval function at inference time.

Our goal is to construct a transformed schema
\vspace{-0.3mm}
\[
\mathcal{\schema}' = g(\mathcal{\schema}, \mathcal{H}),
\]
where $g$ is a schema transformation function. As illustrated in Figure~\ref{fig:pipeline}, we conduct this function $g$ through a two-phase framework. In the Offline Schema Transformation phase, the base schema $\mathcal{\schema}$ and historical workload $\mathcal{H}$ are processed through a modular pool of transformations, namely Schema Abstraction (\texttt{+A}), Schema Partitioning (\texttt{+P}), and Schema Renaming (\texttt{+R}), to produce the optimized schema $\mathcal{\schema}'$. In the Online Prompt Assembly phase, this optimized schema acts as a plug-and-play enhancement that integrates into the flow of any arbitrary Text-to-SQL pipeline $\mathcal{M}$ to process the user question $q$ and generate the query $y$.

\noindent\textbf{Semantic preservation.} Each of the three operator families preserves a different equivalence relation with the base schema. Schema Renaming $g_R$ is a bijection on identifiers, equipped with a mapping dictionary $\mathcal{D}_R$ such that every query over the renamed schema is re-expressible over $\mathcal{\schema}$. Schema Abstraction $g_A$ introduces logical views derivable from $\mathcal{\schema}$ via SQL composition, adding no new information. Schema Partitioning $g_P$ produces a question-specific sub-schema $\mathcal{\schema}^P_q \subseteq \mathcal{\schema}$, restricting but never extending the relations exposed to the model. Therefore, these preservation modes guarantee that $\mathcal{\schema}' = g(\mathcal{\schema}, \mathcal{H})$ never alters the underlying database semantics; it only changes the logical representation presented to the model~\cite{garcia-molina-dbsystems, furst2025evaluating, Halevy2001}. Formal definitions follow in Section~\ref{sec:schema_transformation}.

\begin{Example}
\vspace{-2mm}
Consider the base schema in Figure~\ref{fig:running_example}, with relations
\texttt{molecule}, \texttt{atom}, \texttt{bond}, and \texttt{connected}. For question $Q_1$
\textit{``How many chemical compounds are identified as carcinogenic?''},
the original input to the Text-to-SQL model is the base schema $\mathcal{\schema}$.
Applying renaming $g_R$ yields a schema with descriptive identifiers, e.g., $\texttt{molecule.label} \rightarrow \texttt{Chem\_Molecules.carcinogenic\_flag}$.
Applying partitioning $g_P$ for $Q_1$ then restricts the model-facing schema to $\mathcal{\schema}^P_{Q_1} = \{\texttt{Chem\_Molecules}\}$.
The composition $\mathcal{\schema}' = (g_P \circ g_R)(\mathcal{\schema}, \mathcal{H})$ preserves the underlying database semantics by the modes above; it only changes the logical representation presented to the model.
\vspace{-2mm}
\end{Example}

% \noindent\textbf{Schema Transformation.} Our objective is to learn a transformation function $g: \mathcal{\schema} \to \mathcal{\schema}'$, where $\mathcal{\schema}'$ is an optimized, ``text-to-SQL-friendly'' representation of the original schema. This transformation may involve semantic abstraction, logical entity grouping, or semantic enrichment of table and column identifiers.

\noindent\textbf{Task Objective.} 
Given $\model$, $\workload$, we seek a transformation function $g$ such that the resulting schema $\mathcal{\schema}' = g^*(\mathcal{\schema}, \mathcal{H})$ maximizes downstream Text-to-SQL accuracy:
\begin{equation}
g^*
=
\arg\max_{g}
\mathrm{Acc}\bigl(\model, \workload, g(\mathcal{\schema}, \mathcal{H})\bigr),
\end{equation}
where $\mathrm{Acc}(\model, \workload, \mathcal{\schema}')$ measures the quality of SQL queries generated by $\model$ over workload $\workload$ when conditioned on schema $\mathcal{\schema}' = g(\mathcal{\schema}, \mathcal{H})$, and can be instantiated as
\begin{equation}
\mathrm{Acc}(\model, \workload, \mathcal{\schema}')
=
\frac{1}{|\workload|}
\sum_{q \in \workload}
\mathrm{Score}\bigl(\model(q,\mathcal{\schema}'), y_q^*\bigr),
\end{equation}
where $y_q^*$ is the reference SQL for question $q$, and $\mathrm{Score}$ is an evaluation function such as execution accuracy or semantic equivalence.

Throughout this paper, $g$ is constructed by combining the three operator families introduced above: each configuration applies a subset of $\{g_A, g_P, g_R\}$ jointly, and the seven non-empty combinations are evaluated as distinct configurations in our experiments.
This formulation treats logical schema design itself as an optimization dimension for Text-to-SQL. Rather than only adapting the Text-to-SQL model to the schema, we seek to optimize the schema representation presented to the model.

\subsection{Design Goals and Constraints}

\begin{table}[t]
\centering
\caption{Text-to-SQL-friendly logical design principles.}
%\vspace{-2mm}
\label{tab:design-principles}
\scriptsize
% Added @{} at the ends to remove outer padding (fixes the overspill)
% Tweaked percentages to 0.38, 0.32, 0.22 to give Col 1 more breathing room
\begin{tabular}{@{} >{\raggedright\arraybackslash}p{0.38\linewidth} >{\raggedright\arraybackslash}p{0.32\linewidth} >{\raggedright\arraybackslash}p{0.22\linewidth} @{}}
\toprule
\textbf{Classical design issue} & \textbf{Text-to-SQL bottleneck} & \textbf{Model-facing operator} \\
\midrule
Normalization preserves integrity but fragments data across relations
& Complex join-path reasoning and bridge-table inference
& Schema abstraction (+A) \\
\addlinespace 
% Changed "preserve compactness" to "are compact" to prevent the 3-line wrap
Weak identifiers are compact but obscure data semantics
& Lexical mismatches between the question and schema
& Schema renaming (+R) \\
\addlinespace
Full schemas ensure completeness but enlarge the search space
& Irrelevant context distracts schema linking and retrieval
& Schema partitioning (+P) \\
\bottomrule
\end{tabular}
\vspace{-5mm}
\end{table}

A transformed schema for Text-to-SQL should satisfy three goals: preserve the semantics of the original database, improve the schema's accessibility for natural-language-driven query generation, and remain compatible with existing Text-to-SQL pipelines. These goals, however, are not fully aligned. Improving one aspect of the schema representation may introduce costs in another, making schema design for Text-to-SQL an inherently constrained optimization problem. As summarized in Table~\ref{tab:design-principles}, our framework navigates these constraints by mapping classical design issues to their corresponding Text-to-SQL bottlenecks, and resolving them through three targeted model-facing operators.

\noindent\textbf{Normalization vs.\ reasoning burden.}
Logical schemas used in practice are often normalized, e.g., to Third Normal Form (3NF) or Boyce--Codd Normal Form (BCNF), in order to reduce redundancy and preserve data integrity~\cite{garcia-molina-dbsystems}. While such designs are desirable from a classical database perspective, they often increase the reasoning burden for Text-to-SQL systems. Even relatively simple user questions may require multi-table joins, bridge tables for many-to-many relationships, or nontrivial foreign-key traversal before the intended query structure becomes apparent. 

In the running molecular example, the question about single-bond molecules requires the model to recover an \texttt{atom--bond--molecule} access path before it can perform the final aggregation. The challenge is therefore not to abandon normalization, but to expose higher-level logical abstractions that reduce unnecessary join-path reasoning while preserving query semantics.

\noindent\textbf{Semantic clarity vs.\ lexical ambiguity.}
A second constraint arises from the naming and organization of schema elements. In many real databases, table and column identifiers are abbreviated, overloaded, or only weakly aligned with natural-language concepts. Such lexical ambiguity creates friction for schema linking, since the model must infer the intended meaning of schema elements indirectly. For instance, an identifier such as \texttt{label} in the running example provides little direct evidence that it encodes carcinogenicity.  More descriptive names, enriched metadata, or semantically clearer attribute organization can substantially improve alignment between user questions and schema structure. However, such enrichment must be applied carefully: overly aggressive rewriting may distort established schema conventions or introduce inconsistencies across related schema elements.

\noindent\textbf{Schema completeness vs.\ contextual noise.}
A third trade-off concerns schema scope. Exposing the full database schema ensures that all potentially relevant information is available to the model, but it also enlarges the search space and introduces substantial irrelevant context. This problem becomes especially pronounced in real-world, large, enterprise-level schemas, where most questions touch only a small subset of relations. 
In the running example, the question about connected atoms can be answered from a small schema neighborhood involving \texttt{bond} and \texttt{connected}, but exposing additional relations may encourage unnecessary joins or incorrect projections.
At the same time, excessive pruning risks omitting information required to answer the query~\cite{furst2025evaluating}. A Text-to-SQL-friendly logical design must therefore balance completeness against locality, exposing enough context to support correct query generation while suppressing irrelevant schema regions.

\noindent\textbf{Implications for schema transformation.}
These trade-offs motivate the three operators detailed in Section~\ref{sec:schema_transformation}: schema abstraction shortens rigid join paths; schema partitioning isolates coherent schema neighborhoods; schema renaming improves identifier clarity. Together, they shift part of the reasoning burden from the model's inference process into the logical schema representation itself, constrained by the semantic-preservation modes above.

%\vspace{-2mm} % Now this pulls the header up cleanly
\section{Schema Transformation}
\label{sec:schema_transformation} 

We consider a transformation space that improves Text-to-SQL performance by modifying the logical schema representation exposed to the model while preserving the semantics of the underlying database. In our framework, transformations fall into three broad categories: schema abstraction, schema partitioning with workload-aware organization, and metadata enrichment via schema renaming. These operators may be applied independently or in combination to construct a Text-to-SQL-friendly schema representation.

To illustrate our framework throughout this section, we employ a \textbf{running example} based on a real-world chemistry database. The base schema ($\mathcal{\schema}$) consists of four highly-normalized tables with generic identifiers: 
\begin{itemize}
    \item \texttt{molecule} (\textit{molecule\_id, label})
    \item \texttt{atom} (\textit{atom\_id, molecule\_id, element})
    \item \texttt{bond} (\textit{bond\_id, molecule\_id, bond\_type})
    \item \texttt{connected} (\textit{atom\_id, atom\_id2, bond\_id})
\end{itemize}
\addtolength{\baselineskip}{-0.5pt}
Figure \ref{fig:running_example} visually summarizes how our three transformation techniques operate on this base schema to produce a Text-to-SQL-friendly variant ($\mathcal{\schema}'$).

%\vspace{-1mm}
\subsection{Schema Abstraction}
% A central challenge in Text-to-SQL is the gap between low-level relational structure and high-level user intent. In the real-world schemas that are often normalized, information relevant to a single question is often distributed across multiple relations, requiring the model to infer the appropriate join path before it can perform filtering, aggregation, or projection. As a result, the Text-to-SQL system must solve both a \emph{schema linking} problem and a \emph{query construction} problem.

A central challenge in Text-to-SQL is the gap between low-level relational structure and high-level user intent. Normalized schemas (typically in 3NF or BCNF~\cite{garcia-molina-dbsystems}), such as those in SPIDER~\cite{yu2018spider} and BIRD~\cite{li2023bird} benchmarks, distribute the information needed for a single question across multiple relations, forcing the model to infer the join path before selection, aggregation, or projection in the query. One natural way to reduce this reasoning burden is to expose higher-level logical abstractions through views. In extreme cases, one could imagine an ``omni-view'' that joins every relation and contains all attributes across the schema, thereby eliminating the need for explicit \texttt{JOIN} reasoning. In practice, however, such a representation is neither semantically clean nor operationally reliable; the example below illustrates why it is not the answer.

% One natural way to reduce this burden is to expose higher-level logical abstractions through views. In extreme cases, one could imagine an ``omni-view'' that contains all attributes across the schema, thereby eliminating the need for explicit \texttt{JOIN} reasoning. 
% In practice, however, such a representation is neither semantically clean nor operationally reliable: indiscriminately joining many relations may introduce tuple duplication, spurious combinations, and incorrect aggregation behavior. This is particularly problematic in real-world databases and benchmark schemas such as SPIDER~\cite{yu2018spider} and BIRD~\cite{li2023bird}, which are typically normalized to \textit{Third Normal Form} (3NF) or \textit{Boyce--Codd Normal Form} (BCNF)~\cite{garcia-molina-dbsystems}.
% While normalization improves data integrity and reduces redundancy, it also increases the number of joins needed to answer many user questions. The example below illustrates this hazard. 

\begin{Example}
The ground-truth query $Q_{corr}$ returns $498$ connections for atom $19$. Adding a join through the \texttt{bonds} and \texttt{molecule} tables, as in $Q_{wrong}$, deflates the count to $377$. The cause is that \texttt{INNER JOIN} silently drops every tuple whose join key is \texttt{NULL} on either side---so rows in \texttt{connected} whose \texttt{bond\_id} is \texttt{NULL} (or whose corresponding \texttt{bonds.molecule\_id} is \texttt{NULL}) disappear from the result, even though the selection predicate has not changed.

\vspace{-1mm}
\begin{tcolorbox}[SQLbox, before skip=4pt, after skip=4pt]
\begin{lstlisting}[style=psqlcolor,mathescape,deletendkeywords={YEAR, DATE}]
102 How many connections does the atom 19 have? 
$Q_{corr}=$ SELECT COUNT(T.bond_id) FROM connected AS T 
WHERE SUBSTR(T.atom_id, -2) = '19';

$Q_{wrong}=$ SELECT COUNT(T.bond_id) FROM connected AS T
JOIN bond as b ON b.bond_id = T.bond_id
JOIN molecule  as m on m.molecule_id = b.molecule_id
WHERE SUBSTR(T.atom_id, -2) = '19';

$Q_{corr}$ is 498, $Q_{wrong}$ is 377 because of joining molecule.
\end{lstlisting}
\end{tcolorbox}
\end{Example}
\vspace{-1mm}

Our approach, therefore, synthesizes a set of workload-relevant semantic views that simplify common join patterns without collapsing the schema into a single denormalized table. The objective is to expose higher-level logical access paths that are easier for the model to align with natural-language questions, while retaining sufficient structure to preserve correct query semantics compared to the ``omni-view'' solution. 

Formally, we model schema abstraction as an operator
\setlength{\abovedisplayskip}{4pt}%
\setlength{\belowdisplayskip}{4pt}%
\[
g_A : (\mathcal{\schema}, \mathcal{H}) \rightarrow \mathcal{\schema}^{A},
\]
where $\mathcal{\schema}^{A}$ augments the schema with higher-level logical abstractions, such as derived semantic views, that simplify common access patterns while preserving the semantics of the underlying database.

\begin{Example}
The example below shows the failure that schema abstraction fixes on the single-bond-molecules question: the baseline generates the incorrect query $Q_{wrong}$ that misses the \texttt{molecule} table, while the semantic view \texttt{atom\_join\_bond\_join\_molecule} pre-materializes the join path the question requires, allowing the model to answer the question through a single logical relation.

% The queries below illustrate this motivation. Consider the question: \textit{``How many elements are there for single bond molecules?''} Answering this question over the base schema requires linking the concepts of atoms, bonds, and molecules and inferring the appropriate join path across multiple relations. 
% In practice, the baseline model generates the incorrect query $Q_1$ since it fails to recover the full join structure needed to relate atoms, bonds, and molecules under the intended query semantics. By contrast, the semantic view \texttt{atom\_join\_bond\_join\_molecule} pre-materializes this higher-level access pattern, allowing the model to answer the question through a single logical relation.

%\vspace{-1mm}
\begin{tcolorbox}[SQLbox, breakable, before skip=4pt, after skip=4pt]
\begin{lstlisting}[style=psqlcolor,mathescape,deletendkeywords={YEAR, DATE}]
259 How many elements are there for single bond molecules?
$Q_{corr}=$ SELECT COUNT(DISTINCT element) 
FROM atom_join_bond_join_molecule 
WHERE bond_type = '-';

$Q_{wrong}=$ SELECT COUNT(DISTINCT T1.element) FROM atom AS T1
INNER JOIN bond AS T2 ON T1.molecule_id = T2.molecule_id
WHERE T2.bond_type = '-'
\end{lstlisting}
\end{tcolorbox}
%\vspace{-2mm}
\end{Example}

% This example highlights the role of schema abstraction: the transformed schema allows the model to operate over a single higher-level relation, e.g., \texttt{atom\_join\_bond\_join\_molecule}, thereby avoiding unnecessary join-path inference. More generally, the design challenge is to balance \emph{specificity} and \emph{coverage}. Highly specialized views may be effective for a narrow class of questions, whereas more general views can support a larger portion of the workload. Our goal is therefore to identify a view set that improves downstream Text-to-SQL accuracy without introducing excessive redundancy, semantic distortion, or unnecessary schema complexity.

\vspace{-2mm}
\noindent\textbf{Constraint.} Per the semantic-preservation typing established in Section~\ref{sec:problem_formulation}, views in $\{V_1, \dots, V_n\}$ must be derivable from $\mathcal{\schema}$ via SQL composition, so $g_A$ introduces no new information. The design challenge is to balance \emph{specificity} (a narrow view captures fewer questions) and \emph{coverage} (broader views drift toward the omni-view, risking semantic distortion through indiscriminate joins).

% To resolve this, our methodology synthesizes a set of specialized views tailored to the workload. By creating a semantic view—such as the \texttt{cluster10\_atom\_join\_bond\_join\_molecule} view shown in $Q_2$—the reasoning for finding the correct tables and join paths are skipped entirely. The model now only needs to perform a basic selection and aggregation on a single, unified view. The challenge, therefore, lies in finding the optimal balance between view specificity and coverage. As these views become more generic, their utility across the question set increases. For instance, a view defined as a join between two tables ($T_1 \Join T_2$) can satisfy any question requiring that specific relationship, as well as more complex queries where that join is a subset of the required tables. Our goal is to find the optimal balance between view specificity and coverage to maximize the model's success rate.

\subsection{Schema Partitioning}

Large schemas introduce another difficulty for Text-to-SQL systems: most questions touch only a small subset of the available relations, yet the model is often exposed to the full schema. This creates substantial noise for schema linking and enlarges the search space for query generation. For example, in the BIRD benchmark~\cite{li2023bird}, databases contain an average of roughly 40 tables, whereas a typical question requires access to only a small fraction of them (2.4 on average). This mismatch suggests that much of the schema is irrelevant for any given query and may distract the model from the truly relevant relations ~\cite{furst2025evaluating}.

To address this issue, we first \emph{localize} the relevant partition of the schema for a given question, and then organize it into smaller, semantically coherent partitions. We refine this organization using table co-occurrence patterns and recurring join structures observed in past queries. Each partition may additionally be annotated with auxiliary information such as typical join paths, representative query templates, or salient projection patterns.

%In the zero-shot setting, localization and clustering are derived from schema structure, integrity constraints, and semantic similarity among relations and attributes. When historical workload information is available, we further 

Given a question $q$, we define schema partitioning as the task of selecting a question-specific sub-schema
\[
\mathcal{\schema}^{\mathrm{P}}_q = L(q, \mathcal{\schema}, \mathcal{H}) \subseteq \mathcal{\schema},
\]
where $\mathcal{H}$ denotes historical workload information. The goal is to retain the tables, columns, and join relationships most relevant to answering $q$ while suppressing irrelevant schema partitions.

Beyond reducing schema-level noise during schema linking, the partitioning provides a natural retrieval unit for historical demonstrations: the system first localizes the schema partition for $q$, then retrieves demonstrations from queries assigned to that partition. The example below illustrates the benefit of schema partitioning.

% This localized organization serves two purposes. First, it narrows the candidate schema context presented to the model, thereby reducing irrelevant distractions during schema linking and query generation. Second, in the workload-aware setting, it provides a natural retrieval unit for historical examples. Rather than retrieving demonstrations from the entire workload based on text similarities, the system first localizes the relevant schema region and then retrieves examples associated with that region. In this way, localization reduces schema-level noise, while clustering provides a structured representation for retrieval and reasoning. The example below illustrates the benefit of schema partitioning. 

%\vspace{-2mm}
\begin{Example}
% For the question \textit{``What atoms are connected in single type bonds?''}, a baseline model generates the incorrect four-way join shown in $Q_1$, introducing unnecessary tables and projecting the wrong attributes instead of the connected atom identifiers required by the question. By contrast, our method first localizes the relevant schema region for the question,
On the single-type-bond question below, the baseline produces an unnecessary four-way join in $Q_{wrong}$. In contrast, partitioning the schema to
\[
\mathcal{\schema}^{\mathrm{P}}_q = \{\texttt{bond}, \texttt{connected}\}, \quad \texttt{Partition}_1 = [\texttt{bond}, \texttt{connected}],
\]
and drawing demonstrations from that retrieved partition, $\texttt{Partition}_1$, expose a simpler and more appropriate query pattern, recover the correct projection, leading to the correct query $Q_{corr}$.

% Modern production databases often contain hundreds or thousands of tables, yet the vast majority of user queries target only a small fraction of the total schema. For instance, in the BIRD benchmark \cite{li2023bird}, databases contain an average of 40 tables, while the typical query requires access to only 2.4 tables. This discrepancy suggests that a significant portion of the schema acts as noise, creating an unnecessarily large search space that complicates the LLM's schema-linking and query generation processes.

% To address this, we propose clustering tables into cohesive logical groups to prune the candidates for the LLM. In the absence of historical data, this clustering is performed statically by analyzing schema constraints and semantic similarities. However, when a historical workload is available, we employ a data-driven approach, clustering tables based on their co-occurrence patterns in previous queries. Beyond simple table groupings, each cluster is enriched with auxiliary metadata, including common join paths.

% The advantage of this historical clustering is demonstrated in the example below. When an LLM is asked \textit{"What atoms are connected in single type bonds?"}, modern LLM would generate the massive four-way join seen in $Q_1$ while joining unnecessary tables and projecting the wrong columns.
\vspace{-1mm}
\begin{tcolorbox}[SQLbox, breakable, before skip=4pt, after skip=4pt]
\begin{lstlisting}[style=psqlcolor,mathescape,deletendkeywords={YEAR, DATE}]
210 What atoms are connected in single type bonds?
$Q_{corr}=$ SELECT T2.atom_id, T2.atom_id2 FROM bond AS T1 
INNER JOIN connected AS T2 ON T1.bond_id = T2.bond_id 
WHERE T1.bond_type = '-';

$Q_{wrong}=$ SELECT T3.element, T4.element FROM bond AS T1 
INNER JOIN connected AS T2 ON T1.bond_id = T2.bond_id 
INNER JOIN atom AS T3 ON T2.atom_id = T3.atom_id 
INNER JOIN atom AS T4 ON T2.atom_id2 = T4.atom_id 
WHERE T1.bond_type = '-';

History retrieved from the partitions:
111 Identify all connected atoms with a triple bond.
SELECT T2.atom_id, T2.atom_id2 FROM bond AS T1 
INNER JOIN connected AS T2 ON T1.bond_id = T2.bond_id 
WHERE T1.bond_type = '#'
\end{lstlisting}
\end{tcolorbox}
\vspace{-1mm}
\end{Example}

% 131 What are the atoms that can bond with the atom that has the element lead?
% SELECT T2.atom_id, T2.atom_id2 
% FROM atom AS T1 INNER JOIN connected AS T2 
% ON T1.atom_id = T2.atom_id WHERE T1.element = 'pb'

% The key advantage of this organization is that it supports \emph{localized retrieval}. By restricting retrieval to the most relevant subset of the schema, the model is more likely to observe useful join and projection patterns, and less likely to be distracted by unrelated examples from other schema regions. In this sense, schema partitioning complements schema abstraction: localization identifies the relevant schema neighborhood, while semantic views further simplify access patterns within that neighborhood.
%\vspace{-1mm}
\noindent\textbf{Constraint.} Per Section~\ref{sec:problem_formulation}, $\mathcal{\schema}^{\mathrm{P}}_q$ is a strict subset of $\mathcal{\schema}$ (no new tables introduced), and the partition selection must preserve recall of the relations required to answer $q$. Restricting the demonstration pool to historical queries assigned to the retrieved partitions extends the same locality discipline to retrieval, so the few-shot context shares the partition of the model-facing schema.

\vspace{-1mm}
\subsection{Schema Renaming}

The effectiveness of Text-to-SQL systems depends heavily on the semantic clarity of schema identifiers. In practice, table and column names often vary widely in interpretability, ranging from intuitive descriptors to cryptic abbreviations that require domain knowledge to decode. As observed by prior work such as SNAILS~\cite{snail}, identifier quality has a direct impact on schema understanding and downstream Text-to-SQL performance.

This dependence is particularly important for LLM-based systems, which rely heavily on lexical and semantic alignment between the natural-language question and the schema representation. When identifiers are opaque, overloaded, or weakly descriptive, the model must infer their meaning from context, increasing the risk of schema-linking errors. The following example illustrates this issue. 

\begin{Example}
On the carcinogenic-compounds question below, the generic table name \texttt{molecule} and the ambiguous column identifier \texttt{label = `-'} defeat the baseline (incorrect $Q_{wrong}$); renaming the identifiers to \texttt{Chem\_Molecules.carcinogenic\_flag = `+'} enables the correct query $Q_{corr}$.

% In the base schema, the table name \texttt{molecule} is generic, and the column \texttt{label} is highly ambiguous. Moreover, the value \texttt{'-'} in the \texttt{label} column provides little semantic guidance. 
% As a result, the mapping from the user phrase \textit{``identified as carcinogenic''} to the predicate \texttt{label = '-'} is difficult for the model to infer. The baseline, therefore, generates the incorrect query $Q_1$. After metadata enrichment and schema renaming, the transformed schema exposes the more interpretable relation \texttt{Chem\_Molecules} and the more explicit attribute \texttt{carcinogenic\_flag}, enabling the correct query $Q_2$.

\begin{tcolorbox}[SQLbox, before skip=4pt, after skip=4pt]
\begin{lstlisting}[style=psqlcolor,mathescape,deletendkeywords={YEAR, DATE}]
291 How many chemical compounds in the database are identified as carcinogenic. 
$Q_{corr}=$ SELECT COUNT(*) FROM Chem_Molecules 
WHERE carcinogenic_flag = '+';

$Q_{wrong}=$ SELECT COUNT(`molecule_id`) FROM `molecule` 
WHERE `label` = '-';
\end{lstlisting}
\end{tcolorbox}
\end{Example}
% To improve semantic accessibility, we enrich schema metadata through zero-shot identifier rewriting and description generation. This process produces more descriptive table and column names, resolves avoidable ambiguity, and supplies additional semantic cues that better align the schema with natural-language phrasing. 

Formally, we model metadata enrichment and schema renaming as an operator
\setlength{\abovedisplayskip}{2pt}
\setlength{\belowdisplayskip}{2pt}
\setlength{\abovedisplayshortskip}{2pt}
\setlength{\belowdisplayshortskip}{2pt}
\[
g_R : \mathcal{\schema} \rightarrow \mathcal{\schema}^{R},
\]
where $\mathcal{\schema}^{R}$ preserves the original schema semantics while exposing more descriptive identifiers and metadata.

% Overall, metadata enrichment addresses a different but complementary source of difficulty from the previous two transformations. Whereas schema abstraction reduces structural reasoning burden and schema partitioning reduces schema-level noise, metadata enrichment improves the lexical and semantic alignment between user questions and schema elements. Together, these three transformation classes define the core search space explored by our framework.

\noindent\textbf{Constraint.} According to Section~\ref{sec:problem_formulation}, $g_R$ is bijective on identifiers, and the mapping dictionary $\mathcal{D}_R$ propagates foreign-key constraints onto $\mathcal{\schema}^{R}$. These ensure that every query over $\mathcal{\schema}^{R}$ remains re-expressible over $\mathcal{\schema}$, so the rename is a non-destructive logical layer rather than a structural rewrite of the underlying database.

\section{System Implementation}

\subsection{System Architecture Overview}

\noindent We instantiate the operators $g_A$, $g_P$, and $g_R$ defined in Section~\ref{sec:schema_transformation} as a single middleware layer between the physical database and the Text-to-SQL pipeline. Classic Text-to-SQL pipelines force language models to interact directly with physical database schemas, which are engineered to reduce redundancy rather than for Text-to-SQL friendliness. To bridge this semantic gap, our framework operates as a middleware layer that decouples the physical database schema ($\mathcal{\schema}$) from a highly optimized logical representation ($\mathcal{\schema}'$).

The architecture is divided into two operational phases to shift the computational burden to offline preparation. During the offline phase, the framework analyzes $\mathcal{\schema}$ alongside optional historical query workloads ($\mathcal{H}$) to automatically generate $\mathcal{\schema}^R$ and $\mathcal{\schema}^A$. Specifically, this phase generates the logical layer for Schema Renaming \texttt{(+R)} and Schema Abstraction \texttt{(+A)}, identifies table partitions for Schema Partitioning \texttt{(+P)}, and performs the necessary query rewriting for \texttt{+R} and \texttt{+A}.

During online inference, after obtaining the schema-linking result, the system dynamically retrieves the most relevant partitions and pre-joined views to construct a highly localized prompt context. This decoupled architecture assists the language model by removing the complexities of ambiguous physical identifiers and overwhelming search spaces, enabling highly accurate SQL generation without altering the physical schema.

\subsection{Schema Abstraction Architecture (+A)}
\label{sec:schema_abstraction}
\begin{figure}[t!]
    \centering
    %\vspace{-5pt} % Pulls the diagram up closer to the paragraph above it
    
    % trim = left bottom right top
    % You MUST include 'clip', otherwise the white space is still there, just invisible!
    \includegraphics[page=3, trim=0cm 6.3cm 4.7cm 2.5cm, clip, width=\columnwidth]{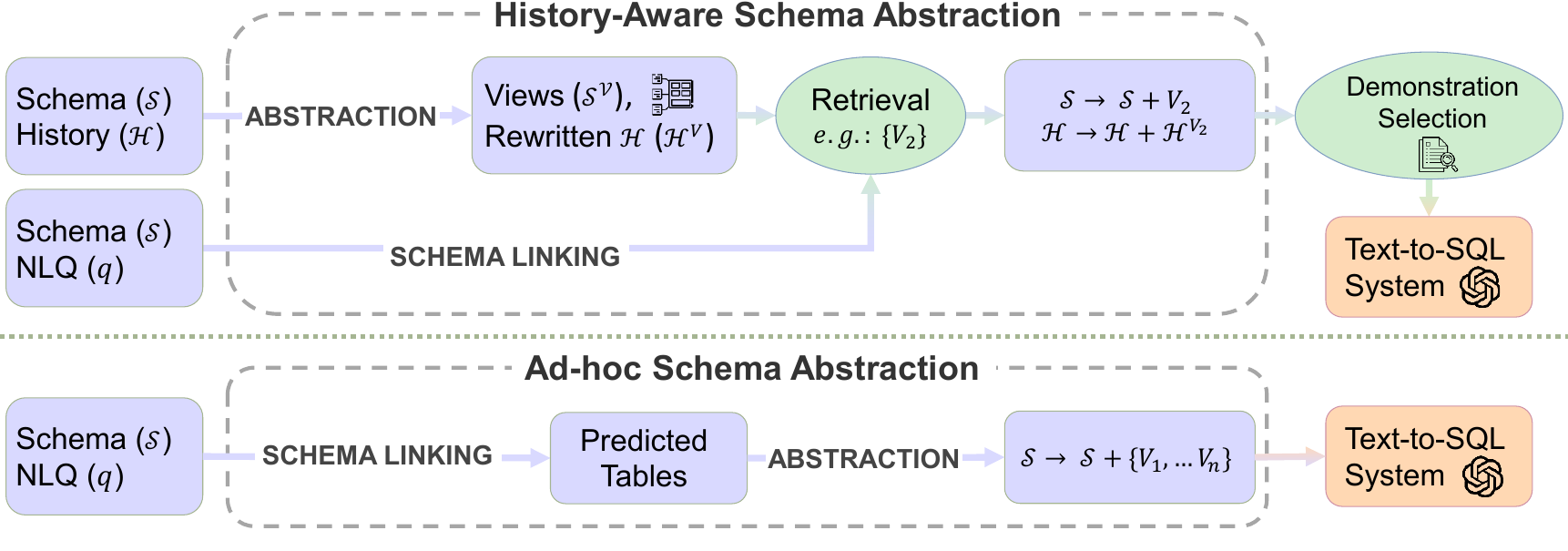}
    
    %\vspace{-5pt} % Snugs the caption slightly closer to the diagram
    \caption{The Schema Abstraction (+A) Architecture.}
    \label{fig:view_synthesis_flow}
    %\vspace{-5pt} % Aggressively pulls the 5.5 heading up closer to the caption
\end{figure}

\noindent The \texttt{+A} implementation proceeds in four steps (illustrated by the dashed bounding box in Figure~\ref{fig:view_synthesis_flow}):
\begin{enumerate}[leftmargin=*]
    \item \textbf{Mine candidate views.} Run frequent-itemset mining in the offline stage over the AST-extracted table sets in $\mathcal{H}$ to obtain recurring multi-table join patterns $\{V_1, \dots, V_n\}$.     
    \item \textbf{Synthesize view DDL.} For each candidate, we prompt the LLM with few-shot demonstrations from $\mathcal{H}$ to generate the join predicates and projections; if the LLM struggles to provide the correct DDL, and when foreign keys are explicit, we fall back on a deterministic code-based generator to enforce the correct execution of the join path. Then, we drop any candidate whose DDL would yield a Cartesian product and cache and persist the rest, forming $\mathcal{\schema}^{A} = \mathcal{\schema} \cup \{V_1, \dots, V_n\}$.
    \item \textbf{Rewrite history.} If the abstracted schema was never used in the few-shot demonstrations, the model might not be able to identify its use. Therefore, we iteratively prompt the LLM to rewrite $\mathcal{H}$ against $\mathcal{\schema}^{A}$ and accept only rewrites that are execution-equivalent to the original, producing $\mathcal{H}^A$.
    \item \textbf{Inject at inference.} For each incoming question, append every view in $\mathcal{\schema}^{A}$ that contains a selected table from schema linking (e.g., $\{V_2\}$ in the figure) to the LLM context, and merge the corresponding $\mathcal{H}^{V_i}$ (queries that access the view $V_i$) into the demonstration pool.
\end{enumerate}

The LLM-driven step~2 has the distinct advantage of inferring undeclared semantic foreign keys (e.g., \texttt{debit\_card\_specializing}), while the code-based fallback enforces deterministic execution when foreign keys are declared. Cached ad-hoc views persist across queries: when a subsequent query requires the same set of tables, the system retrieves the cached view, transitioning ad-hoc synthesis into a persistent optimization strategy. By allowing the LLM to query the augmented $\mathcal{\schema}^{A}$ in step~4, this approach eliminates complex join-key reasoning and reduces the likelihood of join errors.

When $\mathcal{H}$ is empty (no historical workload; zero-shot in prompting), steps~1 and~3 are skipped; steps~2 and~4 instead run on demand at inference, taking the tables identified during schema linking as the input to step~2 (the bottom path of Figure~\ref{fig:view_synthesis_flow}).

\subsection{Schema Partitioning Architecture (+P)}
\label{sec:schema_partitioning}

\begin{figure}[t!]
    \centering
    %\vspace{-5pt} % Pulls the diagram up closer to the paragraph above it
    
    % trim = left bottom right top
    % You MUST include 'clip', otherwise the white space is still there, just invisible!
    \includegraphics[page=2, trim=0cm 10.5cm 4.6cm 2.5cm, clip, width=\columnwidth]{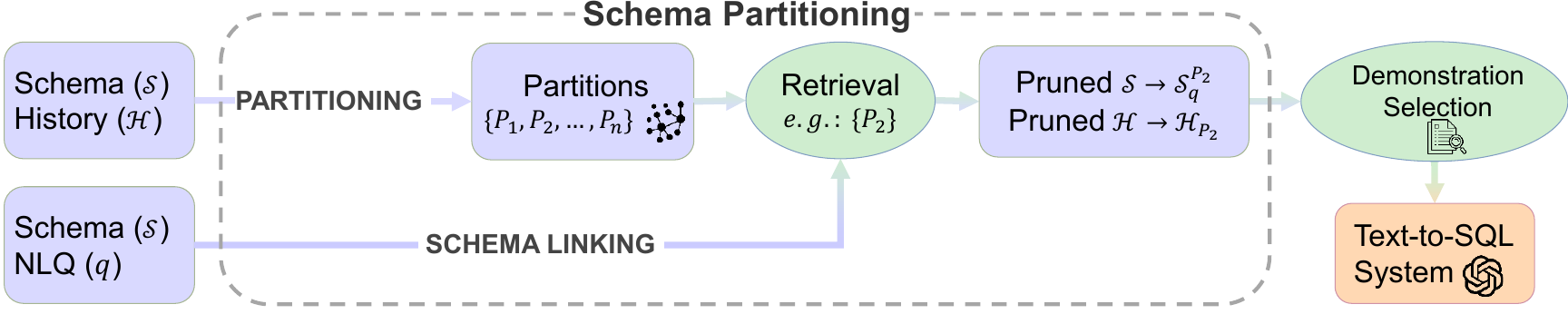}
    
    %\vspace{-5pt} % Snugs the caption slightly closer to the diagram
    \caption{The Schema Partitioning (+P) Architecture.}
    \label{fig:schema_partitioning_flow}
    %\vspace{-5pt} % Reduces space below the figure (helpful if the text reflows later)
\end{figure}

\noindent The \texttt{+P} implementation proceeds in four steps (illustrated by the dashed bounding box in Figure~\ref{fig:schema_partitioning_flow}):
\begin{enumerate}[leftmargin=*]
    \item \textbf{Extract table sets.} Parse the Abstract Syntax Tree (AST) of every query in $\mathcal{H}$ and record the referenced tables.
    \item \textbf{Partitions offline.} Treat the extracted table sets as a frequent-itemset problem; form partitions $\{P_1, \dots, P_n\}$ from table combinations whose frequency exceeds a tunable threshold $\tau$, and assign each query in $\mathcal{H}$, alongside \texttt{Common Join Paths} extracted from $\mathcal{H}$, to the partitions whose tables they reference.
    \item \textbf{Retrieve at inference.} Pass the question $q$ through a schema-linking module to predict the required tables; retrieve any partition containing at least one predicted table (e.g., $\{P_2\}$). 
    \item \textbf{Prune schema and history.} Restrict the model-facing schema to $\mathcal{\schema}^{\mathrm{P}}_q = \bigcup_{P_i \text{ retrieved}} P_i$ (e.g., $\mathcal{\schema}^{\mathrm{P_2}}_q$); restrict the demonstration pool to historical queries assigned to the retrieved partitions (e.g., $\mathcal{H}_{P_2}$); select top-$k$ demonstrations by semantic similarity strictly within this bounded pool.
\end{enumerate}

By treating offline partitioning as a frequent-itemset problem in step~2 rather than relying on static foreign-key topologies or probabilistic graph communities, the system ensures that the partitions represent genuine, frequently utilized access paths grounded in $\mathcal{H}$. The partition-retrieval rule in step~3 maximizes recall while keeping the model-facing schema focused. The hard boundary imposed on the demonstration pool in step~4 constrains the few-shot demonstrations to the exact same partitioned table space as $\mathcal{\schema}^{\mathrm{P}}_q$, providing a focused context with less noise.

\subsection{Schema Renaming Architecture (+R)}

\begin{figure}[t!]
    \centering
    %\vspace{-5pt} % 1. Pulls the diagram slightly closer to the paragraph above it
    
    % trim = left bottom right top
    % You MUST include 'clip', otherwise the white space is still there, just invisible!
    \includegraphics[page=1, trim=5.5cm 7cm 4.5cm 7cm, clip, width=\columnwidth]{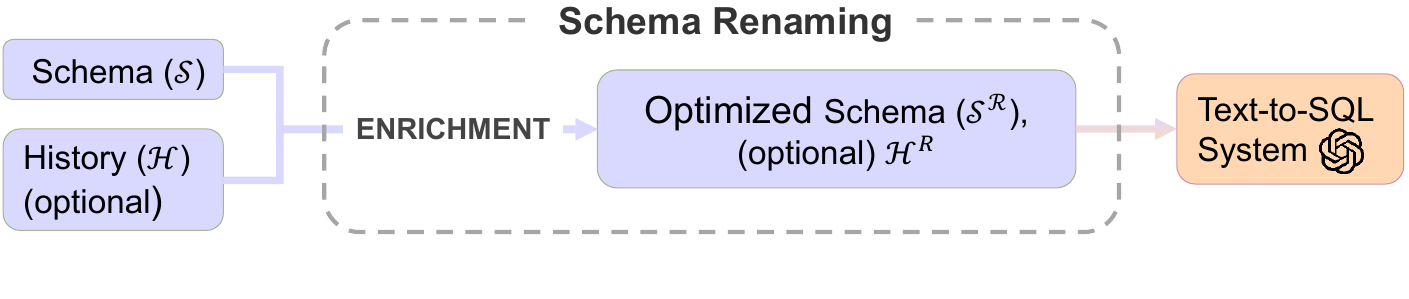}
    
    %\vspace{-5pt} % 2. Snugs the caption slightly closer to the bottom of the diagram
    \caption{The Schema Renaming (+R) Architecture.}
    \label{fig:semantic_enrichment_flow}
    
    %\vspace{-5pt} % 3. Aggressively pulls the 5.3 heading up closer to the caption
\end{figure}

\noindent The \texttt{+R} implementation proceeds in five steps (Figure~\ref{fig:semantic_enrichment_flow}):
\begin{enumerate}[leftmargin=*]
    \item \textbf{Parse the physical schema.} Extract identifiers and foreign-key constraints from the DDL of $\mathcal{\schema}$ into a structured prompt.
    \item \textbf{Generate descriptive aliases.} Prompt the LLM to rename each identifier into a Text-to-SQL-friendly alias; when $\mathcal{H}$ is available, include its query patterns in the prompt to disambiguate cryptic identifiers (e.g., mapping \texttt{A2} to \texttt{district\_name}).
    \item \textbf{Materialize the logical layer.} Instantiate $\mathcal{\schema}^R$ as 1:1 logical views over the base tables (verified by a lightweight integrity check), and construct the mapping dictionary $\mathcal{D}_R: (t_{phys}, c_{phys})$ $\rightarrow (t_{opt}, c_{opt})$ that propagates the base tables' foreign-key constraints onto $\mathcal{\schema}^R$.
    \item \textbf{Rewrite history.} Iteratively prompt the LLM to rewrite $\mathcal{H}$ against $\mathcal{\schema}^R$, accepting only rewrites validated by execution equivalence with the original, producing $\mathcal{H}^R$.
    \item \textbf{Inject at inference.} Provide $\mathcal{\schema}^R$ and $\mathcal{H}^R$ to the LLM, so it operates entirely within the optimized semantic space.
\end{enumerate}

Schema Renaming is non-destructive: the physical schema is untouched and steps~1--4 only add a logical layer on top. Because $\mathcal{\schema}^R$ has a 1:1 mapping over $\mathcal{\schema}$, the SQL generated at step~5 executes directly on the underlying engine without any post-hoc parsing or rewriting. Validating $\mathcal{H}^R$ via execution equivalence in step~4 is critical: $\mathcal{H}^R$ is injected into the prompt at step~5 as few-shot demonstrations, and undetected semantic drift in the rewrites would propagate into the generated SQL.

\subsection{Base Text-to-SQL Pipeline and Online Prompt Assembly}

\begin{figure}[t!]
    \centering
    %\vspace{-5pt}
    \includegraphics[page=4, trim=6cm 5.3cm 9.5cm 1cm, clip, width=\columnwidth]{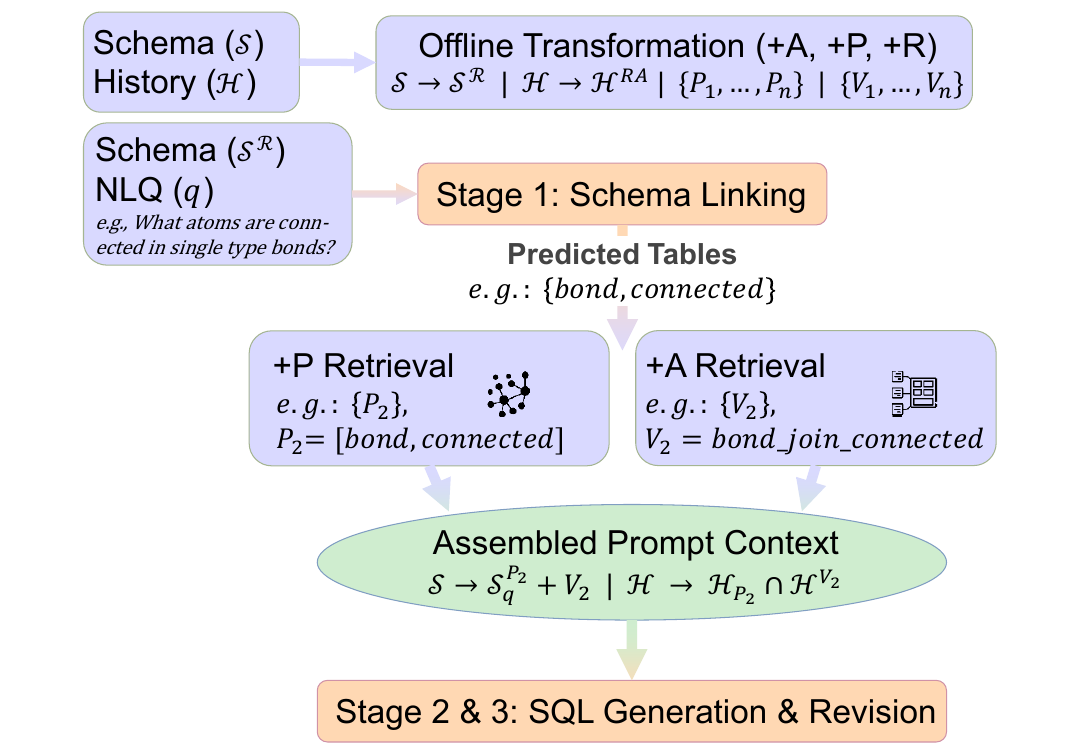}
    %\vspace{-5pt}
    \caption{The online prompt assembly integrating +A+P+R.}
    \label{fig:pipeline_assembly}
    %\vspace{-5pt}
\end{figure}

\subsubsection{Base Text-to-SQL Pipeline (\mypipeline{})}
\noindent To evaluate and isolate the impact of logical schema optimization from complex prompt engineering heuristics, we evaluate our framework using a streamlined three-stage baseline pipeline, \mypipeline{}. Stage 1 performs initial schema linking (inspired by DIN-SQL~\cite{pourreza2023dinsql}). Stages 2 and 3 handle SQL generation and revision using CHESS~\cite{talaei2025chess} prompts, leveraging Stage 2's query's runtime diagnostics for Stage 3 to correct structural flaws and produce the final SQL query.

\subsubsection{Full Pipeline Prompt Assembly (+A+P+R)}
\noindent Our framework integrates seamlessly into the \mypipeline{} baseline (or any Text-to-SQL pipeline) by dynamically assembling an optimized schema and prompt context along the standard flow. As illustrated in Figure~\ref{fig:pipeline_assembly}, this integration occurs across three assembly points:

\noindent\textbf{1. Offline Transformation (+A, +P, +R):} Computed entirely offline before any query is processed, \texttt{+R} transforms the physical schema $\mathcal{\schema}$ into the optimized schema $\mathcal{\schema}^R$ and rewrites the historical workload $\mathcal{H}$ into $\mathcal{H}^R$. Utilizing $\mathcal{\schema}^R$ and $\mathcal{H}^R$, the offline phases of \texttt{+P} and \texttt{+A} generate the partitions $\{P_1, \dots, P_n\}$(e.g., the partition with \texttt{[bond, connected]}), logical views $\{V_1, \dots, V_n\}$ (e.g., \texttt{bond\_join\_connected}), and rewritten history $\mathcal{H}^{RA}$.    

\noindent\textbf{2. Dynamic Schema Assembly (+A, +P):} Between stage 1 and 2, the predicted tables trigger the retrieval modules. \texttt{+P} retrieves relevant partitions (e.g., $P_2$), pruning the schema down to $\mathcal{\schema}^{\mathrm{P}}_q$(e.g., $\mathcal{\schema}^{\mathrm{P_2}}_q$) and restricting the historical pool to $\mathcal{H}_{P_2}$ for top-$k$ retrieval. Concurrently, \texttt{+A} appends relevant logical views (e.g., $V_2$) and their corresponding history (e.g., $\mathcal{H}^{V_2}$) to the historical pool.

\noindent\textbf{3. Stage 2 \& 3: SQL Generation and Revision:} The final assembled context, comprising the unified schema ($\mathcal{\schema}^{\mathrm{P_2}}_q + V_2$) and the focused top-$k$ history ($\mathcal{H}_{P_2} \cap \mathcal{H}^{V_2}$), is provided to the LLM. Both generation and revision operate within this identical optimized semantic space, with Stage 3 additionally incorporating the runtime execution feedback of the generated query from stage 2.
\section{Experiments}

We evaluate whether logical schema transformation improves Text-to-SQL performance across benchmarks, pipelines, LLM backbones, and workload history availability settings. Our evaluation is designed to answer the following research questions:

\begin{itemize}[leftmargin=2em]
    \item \textbf{RQ1:} Does logical schema transformation improve Text-to-SQL performance when historical workloads are available?
    \item \textbf{RQ2:} How do different transformation operators behave across pipelines and benchmarks?
    \item \textbf{RQ3:} Are the gains robust across LLM backbones and in zero-shot settings without historical workloads?
    \item \textbf{RQ4:} When does schema transformation help, and when can it hurt?
\end{itemize}

\subsection{Experimental Setup}
\begin{table*}[t]
    \centering
    \caption{
    History-aware results on BIRD-Union across Text-to-SQL pipelines using top-$k{=}3$ retrieved demonstrations on GPT-4.1 mini unless noted. Values above the baseline are \textit{italicized}; the best score per approach is bolded.
    % Performance comparison with History SQL for BIRD-Union. Values higher than the baseline are \textit{italicized}, and the highest score per approach is \textbf{bolded}. (A: Abstraction, P: Partitioning, R: Rename) - topk = 3
    }
    \label{tab:bird-union-with-history}
    % \resizebox{\linewidth}{!}{%
    {
    \begin{tabular}{l cccccccc}
        \toprule
        \textbf{Approach} & \textbf{Baseline} & \textbf{+A} & \textbf{+P} & \textbf{+R} & \textbf{+A+P} & \textbf{+A+R} & \textbf{+P+R} & \textbf{+A+P+R} \\
        \midrule
        \mypipeline{} & 49.8 & \textit{50.9}~\textcolor{green!60!black}{\scriptsize (+1.1)} & \textit{50.5}~\textcolor{green!60!black}{\scriptsize (+0.7)} & \textit{50.5}~\textcolor{green!60!black}{\scriptsize (+0.7)} & \textbf{\textit{52.4}}~\textcolor{green!60!black}{\scriptsize (+2.6)} & \textit{51.4}~\textcolor{green!60!black}{\scriptsize (+1.6)} & \textit{50.5}~\textcolor{green!60!black}{\scriptsize (+0.7)} & \textit{52.3}~\textcolor{green!60!black}{\scriptsize (+2.5)} \\
        \mypipeline{} (50\% Hist.) & 48.8 & \textit{50.5}~\textcolor{green!60!black}{\scriptsize (+1.7)} & \textit{51.0}~\textcolor{green!60!black}{\scriptsize (+2.2)} & \textit{49.9}~\textcolor{green!60!black}{\scriptsize (+1.1)} & \textit{50.3}~\textcolor{green!60!black}{\scriptsize (+1.5)} & \textit{50.6}~\textcolor{green!60!black}{\scriptsize (+1.8)} & \textit{51.0}~\textcolor{green!60!black}{\scriptsize (+2.2)} & \textbf{\textit{51.9}}~\textcolor{green!60!black}{\scriptsize (+3.1)} \\
        % \mypipeline{} (Gemini 2.5-flash) & 48.1 & 48.0~\textcolor{gray}{\scriptsize (-0.1)} & \textit{50.7}~\textcolor{green!60!black}{\scriptsize (+2.6)} & \textit{48.6}~\textcolor{green!60!black}{\scriptsize (+0.5)} & \textit{51.6}~\textcolor{green!60!black}{\scriptsize (+3.5)} & \textit{50.6}~\textcolor{green!60!black}{\scriptsize (+2.5)} & \textit{51.4}~\textcolor{green!60!black}{\scriptsize (+3.3)} & \textbf{\textit{52.3}}~\textcolor{green!60!black}{\scriptsize (+4.2)} \\
        % \mypipeline{} (GPT-5.4 mini) & 52.8 & \textbf{\textit{55.5}}~\textcolor{green!60!black}{\scriptsize (+2.7)} & \textit{53.1}~\textcolor{green!60!black}{\scriptsize (+0.3)} & \textit{52.9}~\textcolor{gray}{\scriptsize (+0.1)} & \textit{54.5}~\textcolor{green!60!black}{\scriptsize (+1.7)} & \textit{54.5}~\textcolor{green!60!black}{\scriptsize (+1.7)} & \textit{54.0}~\textcolor{green!60!black}{\scriptsize (+1.2)} & \textit{53.9}~\textcolor{green!60!black}{\scriptsize (+1.1)} \\
        DIN-SQL & 44.2 & \textit{45.2}~\textcolor{green!60!black}{\scriptsize (+1.0)} & \textit{48.1}~\textcolor{green!60!black}{\scriptsize (+3.9)} & \textit{46.7}~\textcolor{green!60!black}{\scriptsize (+2.5)} & \textit{46.7}~\textcolor{green!60!black}{\scriptsize (+2.5)} & \textit{45.9}~\textcolor{green!60!black}{\scriptsize (+1.7)} & \textbf{\textit{48.2}}~\textcolor{green!60!black}{\scriptsize (+4.0)} & \textbf{\textit{48.2}}~\textcolor{green!60!black}{\scriptsize (+4.0)} \\
        MAC-SQL & 45.9 & \textit{47.2}~\textcolor{green!60!black}{\scriptsize (+1.3)} & \textit{48.0}~\textcolor{green!60!black}{\scriptsize (+2.1)} & \textit{46.4}~\textcolor{green!60!black}{\scriptsize (+0.5)} & \textit{48.5}~\textcolor{green!60!black}{\scriptsize (+2.6)} & \textit{48.5}~\textcolor{green!60!black}{\scriptsize (+2.6)} & \textit{48.4}~\textcolor{green!60!black}{\scriptsize (+2.5)} & \textbf{\textit{48.6}}~\textcolor{green!60!black}{\scriptsize (+2.7)} \\
        CSCSQL (Qwen2.5-Coder-7B)                & 49.0 & \textit{49.4}~\textcolor{green!60!black}{\scriptsize (+0.4)} & \textbf{\textit{50.6}}~\textcolor{green!60!black}{\scriptsize (+1.6)} & \textit{50.3}~\textcolor{green!60!black}{\scriptsize (+1.3)} & 49.0~\textcolor{gray}{\scriptsize (+0.0)} & \textit{50.1}~\textcolor{green!60!black}{\scriptsize (+1.1)} & \textit{49.9}~\textcolor{green!60!black}{\scriptsize (+0.9)} & \textit{49.7}~\textcolor{green!60!black}{\scriptsize (+0.7)} \\
        \bottomrule
    \end{tabular}%
    }
\end{table*}

\begin{table*}[t]
    \centering
    \caption{
    History-aware results on Spider-Union across Text-to-SQL pipelines using top-$k{=}3$ retrieved demonstrations on GPT-4.1 mini unless noted. 
    Values above the baseline are \textit{italicized}; the best score per approach is bolded.
    % Performance comparison with History SQL for Spider-Union. Values higher than the baseline are \textit{italicized}, and the highest score per approach is \textbf{bolded}. (A: Abstraction, P: Partitioning, R: Rename) - topk = 3
    }
    \label{tab:spider-union-with-history}
    % \resizebox{\linewidth}{!}{%
    {
    \begin{tabular}{l cccccccc}
        \toprule
        \textbf{Approach} & \textbf{Baseline} & \textbf{+A} & \textbf{+P} & \textbf{+R} & \textbf{+A+P} & \textbf{+A+R} & \textbf{+P+R} & \textbf{+A+P+R} \\
        \midrule
        \mypipeline{} & 79.5 & \textbf{\textit{81.3}}~\textcolor{green!60!black}{\scriptsize (+1.8)} & \textit{79.5}~\textcolor{gray}{\scriptsize (0.0)} & \textit{80.1}~\textcolor{green!60!black}{\scriptsize (+0.6)} & \textit{79.9}~\textcolor{green!60!black}{\scriptsize (+0.4)} & \textit{79.7}~\textcolor{green!60!black}{\scriptsize (+0.2)} & \textit{79.9}~\textcolor{green!60!black}{\scriptsize (+0.4)} & \textbf{\textit{81.3}}~\textcolor{green!60!black}{\scriptsize (+1.8)} \\
        \mypipeline{} (50\% Hist.) & 77.9 & \textbf{\textit{79.5}}~\textcolor{green!60!black}{\scriptsize (+1.6)} & 77.1~\textcolor{red}{\scriptsize (-0.8)} & \textit{78.1}~\textcolor{green!60!black}{\scriptsize (+0.2)} & \textit{78.1}~\textcolor{green!60!black}{\scriptsize (+0.2)} & \textit{78.5}~\textcolor{green!60!black}{\scriptsize (+0.6)} & \textit{78.9}~\textcolor{green!60!black}{\scriptsize (+1.0)} & \textit{79.3}~\textcolor{green!60!black}{\scriptsize (+1.4)} \\
        % \mypipeline{} (Gemini 2.5-flash-lite) & 80.5 & 80.5~\textcolor{gray}{\scriptsize (0.0)} & \textit{81.7}~\textcolor{green!60!black}{\scriptsize (+1.2)} & \textit{81.5}~\textcolor{green!60!black}{\scriptsize (+1.0)} & \textit{81.5}~\textcolor{green!60!black}{\scriptsize (+1.0)} & \textit{81.3}~\textcolor{green!60!black}{\scriptsize (+0.8)} & \textit{82.9}~\textcolor{green!60!black}{\scriptsize (+2.4)} & \textbf{\textit{83.1}}~\textcolor{green!60!black}{\scriptsize (+2.6)} \\
        % \mypipeline{} (GPT-5.4 mini) & 81.7 & \textbf{\textit{84.5}}~\textcolor{green!60!black}{\scriptsize (+2.8)} & 82.5~\textcolor{green!60!black}{\scriptsize (+0.8)} & \textit{82.5}~\textcolor{green!60!black}{\scriptsize (+0.8)} & \textit{82.3}~\textcolor{green!60!black}{\scriptsize (+0.6)} & \textit{84.1}~\textcolor{green!60!black}{\scriptsize (+2.4)} & \textbf{\textit{84.5}}~\textcolor{green!60!black}{\scriptsize (+2.8)} & \textit{84.3}~\textcolor{green!60!black}{\scriptsize (+2.6)} \\
        DIN-SQL & 74.9 & \textit{76.3}~\textcolor{green!60!black}{\scriptsize (+1.4)} & \textit{75.3}~\textcolor{green!60!black}{\scriptsize (+0.4)} & \textit{77.5}~\textcolor{green!60!black}{\scriptsize (+2.6)} & \textit{76.5}~\textcolor{green!60!black}{\scriptsize (+1.6)} & \textit{75.7}~\textcolor{green!60!black}{\scriptsize (+0.8)} & \textbf{\textit{79.1}}~\textcolor{green!60!black}{\scriptsize (+4.2)} & \textit{77.5}~\textcolor{green!60!black}{\scriptsize (+2.6)} \\
        MAC-SQL & 76.5 & \textit{77.5}~\textcolor{green!60!black}{\scriptsize (+1.0)} & \textit{77.3}~\textcolor{green!60!black}{\scriptsize (+0.8)} & \textit{78.9}~\textcolor{green!60!black}{\scriptsize (+2.4)} & \textit{78.9}~\textcolor{green!60!black}{\scriptsize (+2.4)} & \textbf{\textit{79.7}}~\textcolor{green!60!black}{\scriptsize (+3.2)} & \textit{77.9}~\textcolor{green!60!black}{\scriptsize (+1.4)} & \textit{78.1}~\textcolor{green!60!black}{\scriptsize (+1.6)} \\
        CSCSQL (Qwen2.5-Coder-7B)                & 82.7 & \textit{84.1}~\textcolor{green!60!black}{\scriptsize (+1.4)} & 83.1~\textcolor{green!60!black}{\scriptsize (+0.4)} & \textbf{\textit{84.5}}~\textcolor{green!60!black}{\scriptsize (+1.8)} & \textit{84.1}~\textcolor{green!60!black}{\scriptsize (+1.4)} & \textit{83.9}~\textcolor{green!60!black}{\scriptsize (+1.2)} & \textit{83.1}~\textcolor{green!60!black}{\scriptsize (+0.4)} & \textbf{\textit{84.5}}~\textcolor{green!60!black}{\scriptsize (+1.8)} \\
        \bottomrule
    \end{tabular}%
    }
\end{table*}

\subsubsection{Benchmarks}
We evaluate on two union-style Text-to-SQL benchmarks, BIRD-Union and Spider-Union, constructed from the development sets of BIRD~\cite{li2023bird} and Spider~\cite{yu2018spider}, respectively. These benchmarks are designed to stress schema understanding by exposing the model to larger and less neatly separated schema spaces than the original per-database evaluation setting.

\begin{itemize}[leftmargin=2em]
    \item \textbf{BIRD-Union.}
    BIRD-Union combines the 11 databases from the BIRD~\cite{li2023bird} development set into a single unified schema. The resulting benchmark contains 75 tables and 1,534 Text-to-SQL pairs. We omit evidence and column descriptions from the prompt, so the model must rely primarily on the natural-language question, the schema, and optionally historical workload information. This setting is challenging because the unified schema contains semantically diverse domains, ambiguous identifiers, and a substantially larger schema search space.

    \item \textbf{Spider-Union.}
    Spider-Union combines 19 databases from the development set of the Spider benchmark~\cite{yu2018spider} into a single unified schema. The resulting benchmark contains 78 tables and 1,004 Text-to-SQL pairs. Compared with BIRD-Union, Spider-Union generally contains cleaner schema identifiers and less external domain knowledge, but still stresses schema linking and table selection due to the enlarged schema context.
\end{itemize}

Overall, BIRD-Union is the more challenging benchmark because it combines more complex questions, weaker schema semantics, and the absence of evidence and column descriptions. Spider-Union provides a complementary setting for testing whether logical schema transformation remains useful on a benchmark with a cleaner schema design and less domain-specific evidence.

\subsubsection{Pipelines and backbones}
We evaluate our framework with multiple Text-to-SQL pipelines, including \mypipeline{}, DIN-SQL~\cite{pourreza2023dinsql}, MAC-SQL~\cite{wang2025mac}, and CSCSQL~\cite{CSCSQL}. For API-based pipelines, we use the same proprietary LLM backbones and decoding configuration across the baseline and transformed-schema variants to ensure a controlled comparison. For local-model experiments, we evaluate CSCSQL variants based on Qwen2.5-Coder and run inference on a single NVIDIA A40 GPU using FP16 precision.

\subsubsection{Schema transformations}
We evaluate three classes of logical schema transformations: \textbf{A} for schema abstraction, \textbf{P} for schema partitioning, and \textbf{R} for schema renaming. In zero-shot settings, the framework applies the schema-only \textbf{R} transformation, with \textbf{A} falling back to an ad-hoc variant of schema abstraction (Section~\ref{sec:schema_abstraction}). In history-aware settings, all three operators consume the historical workload $\mathcal{H}$ to produce workload-aware variants, with \textbf{P} additionally supporting partition retrieval and schema organization. We evaluate each operator individually and in combination.

\subsubsection{Metric}
We report execution accuracy (EX), the standard metric for Text-to-SQL evaluation. For each pipeline and benchmark, we compare the original baseline against transformed-schema variants. Improvements are reported as absolute percentage-point gains over the corresponding baseline.

\subsubsection{Historical workload setting.}
For history-aware experiments, we partition the Text-to-SQL pairs into two sets: a historical workload and a held-out test workload. Unless otherwise stated, 50\% of the questions are assigned to the historical workload, and the remaining 50\% are used for evaluation. The historical workload is used in two ways: first, to guide workload-aware schema transformations such as partitioning and abstraction; second, to retrieve prompt-time demonstrations for few-shot SQL generation. 
The baseline pipelines \emph{also draw their few-shot demonstrations from the historical workload}, so the only difference between baseline and transformed-schema variants is the schema representation itself. Test questions are \emph{excluded from the historical workload} to prevent data leakage while evaluating whether workload-aware schema organization and retrieval improve downstream generation.

We distinguish between \emph{design-time history} and \emph{prompt-time history}. Design-time history refers to historical workload information available to the schema optimization module during the construction of transformed schemas. Prompt-time history refers to the retrieved question-SQL demonstrations in the Text-to-SQL prompt. While both refer to the same historical pool, this distinction allows us to evaluate whether gains come from the transformed schema itself, from few-shot demonstrations, or their combinations.

We also conduct an ablation study on historical workload availability by varying the amount of history available to the system, using 0\%, 50\%, and 100\% of the historical pool (note that few-shot prompting is not available in the 0\% history setting). The test set remains fixed across these settings.

%\vspace{-2mm}
\subsection{Main Results With Historical Workloads}
\begin{table*}[t]
    \centering
    \caption{
    Backbone robustness of \textsc{BaseSQL} on BIRD-Union and Spider-Union under the history-aware setting with the same schema transformation on various LLMs. Values above the baseline are \textit{italicized}; and the best score per backbone is bolded.
    }
    %\vspace{-3mm}
    \label{tab:bird-union-with-history-backbone}
    \resizebox{\linewidth}{!}{%
    % \small
    % {
    \begin{tabular}{l l cccccccc}
        \toprule
        \textbf{Approach} & \textbf{Dataset} & \textbf{Baseline} & \textbf{+A} & \textbf{+P} & \textbf{+R} & \textbf{+A+P} & \textbf{+A+R} & \textbf{+P+R} & \textbf{+A+P+R} \\
        \midrule
        \mypipeline{} (GPT-4.1 mini) & BIRD-UNION & 49.8 & \textit{50.9}~\textcolor{green!60!black}{\scriptsize (+1.1)} & \textit{50.5}~\textcolor{green!60!black}{\scriptsize (+0.7)} & \textit{50.5}~\textcolor{green!60!black}{\scriptsize (+0.7)} & \textbf{\textit{52.4}}~\textcolor{green!60!black}{\scriptsize (+2.6)} & \textit{51.4}~\textcolor{green!60!black}{\scriptsize (+1.6)} & \textit{50.5}~\textcolor{green!60!black}{\scriptsize (+0.7)} & \textit{52.3}~\textcolor{green!60!black}{\scriptsize (+2.5)} \\
        \mypipeline{} (Gemini 2.5-flash-lite)  & BIRD-UNION & 48.1 & 48.0~\textcolor{gray}{\scriptsize (-0.1)} & \textit{50.7}~\textcolor{green!60!black}{\scriptsize (+2.6)} & \textit{48.6}~\textcolor{green!60!black}{\scriptsize (+0.5)} & \textit{51.6}~\textcolor{green!60!black}{\scriptsize (+3.5)} & \textit{50.6}~\textcolor{green!60!black}{\scriptsize (+2.5)} & \textit{51.4}~\textcolor{green!60!black}{\scriptsize (+3.3)} & \textbf{\textit{52.3}}~\textcolor{green!60!black}{\scriptsize (+4.2)} \\
        \mypipeline{} (GPT-5.4 mini)  & BIRD-UNION & 52.8 & \textbf{\textit{55.5}}~\textcolor{green!60!black}{\scriptsize (+2.7)} & \textit{53.1}~\textcolor{green!60!black}{\scriptsize (+0.3)} & \textit{52.9}~\textcolor{gray}{\scriptsize (+0.1)} & \textit{54.5}~\textcolor{green!60!black}{\scriptsize (+1.7)} & \textit{54.5}~\textcolor{green!60!black}{\scriptsize (+1.7)} & \textit{54.0}~\textcolor{green!60!black}{\scriptsize (+1.2)} & \textit{53.9}~\textcolor{green!60!black}{\scriptsize (+1.1)} \\\hline

        \mypipeline{} (GPT-4.1 mini) & Spider-UNION & 79.5 & \textbf{\textit{81.3}}~\textcolor{green!60!black}{\scriptsize (+1.8)} & 79.5~\textcolor{gray}{\scriptsize (0.0)} & \textit{80.1}~\textcolor{green!60!black}{\scriptsize (+0.6)} & \textit{79.9}~\textcolor{green!60!black}{\scriptsize (+0.4)} & \textit{79.7}~\textcolor{green!60!black}{\scriptsize (+0.2)} & \textit{79.9}~\textcolor{green!60!black}{\scriptsize (+0.4)} & \textbf{\textit{81.3}}~\textcolor{green!60!black}{\scriptsize (+1.8)} \\
        \mypipeline{} (Gemini 2.5-flash-lite) & Spider-UNION & 80.5 & 80.5~\textcolor{gray}{\scriptsize (0.0)} & \textit{81.7}~\textcolor{green!60!black}{\scriptsize (+1.2)} & \textit{81.5}~\textcolor{green!60!black}{\scriptsize (+1.0)} & \textit{81.5}~\textcolor{green!60!black}{\scriptsize (+1.0)} & \textit{81.3}~\textcolor{green!60!black}{\scriptsize (+0.8)} & \textit{82.9}~\textcolor{green!60!black}{\scriptsize (+2.4)} & \textbf{\textit{83.1}}~\textcolor{green!60!black}{\scriptsize (+2.6)} \\
        \mypipeline{} (GPT-5.4 mini) & Spider-UNION & 81.7 & \textbf{\textit{84.5}}~\textcolor{green!60!black}{\scriptsize (+2.8)} & \textit{82.5}~\textcolor{green!60!black}{\scriptsize (+0.8)} & \textit{82.5}~\textcolor{green!60!black}{\scriptsize (+0.8)} & \textit{82.3}~\textcolor{green!60!black}{\scriptsize (+0.6)} & \textit{84.1}~\textcolor{green!60!black}{\scriptsize (+2.4)} & \textbf{\textit{84.5}}~\textcolor{green!60!black}{\scriptsize (+2.8)} & \textit{84.3}~\textcolor{green!60!black}{\scriptsize (+2.6)} \\
        \bottomrule
    \end{tabular}%
    }
    %\vspace{-3mm}
\end{table*}

Tables~\ref{tab:bird-union-with-history} and~\ref{tab:spider-union-with-history} report the main results on BIRD-Union and Spider-Union under the history-aware setting. \textbf{Across both benchmarks, every pipeline improves under at least six of seven configurations.}
The remainder of this section unpacks how, where, and when these gains arise: Section~\ref{sec:transformation-types} examines which operators drive the gains; Section~\ref{sec:robustness} tests robustness across backbones and workload-history availability; Section~\ref{sec:per-domain-analysis} dissects where the improvements concentrate and when transformations can regress.

\noindent\textbf{BIRD-Union.} 
On BIRD-Union, logical schema transformation improves performance broadly across pipelines and configurations. Every evaluated system improves under at least six of the seven transformed-schema configurations. 
% For \mypipeline{}, performance improves from 49.8 to 52.4 with GPT-4.1 mini, and from 48.1 to 52.3 with Gemini 2.5-flash-lite. The gains also transfer to existing Text-to-SQL pipelines:  DIN-SQL~\cite{pourreza2023dinsql} improves from 44.2 to 48.2, MAC-SQL~\cite{wang2025mac} from 45.9 to 48.6, and CSCSQL~\cite{CSCSQL} from 49.0 to 50.6. These improvements are meaningful because the baselines already exploit historical workload information in the few-shot prompt.  Thus, the results suggest that transforming the schema representation can provide complementary gains even when the model has access to relevant past question-SQL pairs.
For \mypipeline{} with GPT-4.1 mini, performance improves from 49.8 to 52.4 (best under \texttt{+A+P}). The gains also transfer to existing Text-to-SQL pipelines: DIN-SQL~\cite{pourreza2023dinsql} improves from 44.2 to 48.2 (best under \texttt{+P+R} or \texttt{+A+P+R}), MAC-SQL~\cite{wang2025mac} improves from 45.9 to 48.6 (best under \texttt{+A+P+R}), and CSCSQL~\cite{CSCSQL} improves from 49.0 to 50.6 (best under \texttt{+P}). These improvements are meaningful because the baselines already exploit historical workload information in the few-shot prompt: transforming the schema representation provides complementary gains even when the model has access to relevant past question-SQL pairs.

\noindent\textbf{Spider-Union.}
% A similar pattern holds for Spider-Union. Every evaluated system improves under most transformation settings. \mypipeline{} improves from 79.5 to 81.3 with GPT-4.1 mini, from 80.5 to 83.1 with Gemini 2.5-flash-lite, and from 81.7 to 84.5 with GPT-5.4 mini. DIN-SQL improves from 74.9 to 79.1, MAC-SQL improves from 76.5 to 79.7, and CSCSQL improves from 82.7 to 84.5. These results show that the benefit of logical schema transformation is not tied to a single model family, model scale, or pipeline design.
A similar pattern holds for Spider-Union. With GPT-4.1 mini, \mypipeline{} improves from 79.5 to 81.3 (best under \texttt{+A} or \texttt{+A+P+R}), DIN-SQL improves from 74.9 to 79.1 (best under \texttt{+P+R}), MAC-SQL improves from 76.5 to 79.7 (best under \texttt{+A+R}), and CSCSQL improves from 82.7 to 84.5 (best under \texttt{+R} or \texttt{+A+P+R}). The benefit of logical schema transformation, therefore, extends to a cleaner schema benchmark and persists across pipeline designs.

Across BIRD-Union and Spider-Union, every evaluated pipeline improves under most transformation configurations, and all pipelines obtain their best results from a transformed schema.
Combined transformations often perform better than individual operators, indicating that different schema-related bottlenecks can interact. However, the full combination \texttt{+A+P+R} is not uniformly optimal: some pipelines benefit most from renaming, others from schema abstraction, and others from smaller combinations such as \texttt{+P+R} or \texttt{+A+P}. This heterogeneity suggests that logical schema design should be treated as an adaptive optimization space, where the appropriate transformation depends on the downstream pipeline, workload history, and dominant schema-related failure mode.

\subsection{Analysis of Transformation Types}
\label{sec:transformation-types}

% Across the main result tables (Table~\ref{tab:bird-union-with-history}, Table~\ref{tab:spider-union-with-history}), the impact of individual transformations is clearly heterogeneous.
% \textbf{The best-performing transformation is heterogeneous across pipelines and benchmarks: no single combination dominates.} On BIRD-Union the best configuration is \texttt{+P+A} for \mypipeline{}, \texttt{+R+P} (or equivalently \texttt{+R+A+P}) for DIN-SQL, \texttt{+R+A+P} for MAC-SQL, and \texttt{+P} for CSCSQL; on Spider-Union the best switches to \texttt{+A} for \mypipeline{}, \texttt{+R+P} for DIN-SQL, \texttt{+R+A} for MAC-SQL, and \texttt{+R} (or \texttt{+R+A+P}) for CSCSQL. Combined transformations often beat individual operators, but \texttt{+R+A+P} is not uniformly optimal. This heterogeneity reinforces our framing: logical schema design is an adaptive optimization space whose right configuration depends on the dominant schema-related failure mode of the pipeline.
\textbf{Multiple transformation configurations are competitive on each pipeline; the strongest combination varies modestly across pipelines and benchmarks.} On BIRD-Union the best is \texttt{+A+P} for \mypipeline{}, \texttt{+A+P+R} (tied with \texttt{+P+R}) for DIN-SQL, \texttt{+A+P+R} for MAC-SQL, and \texttt{+P} for CSCSQL; on Spider-Union the best is \texttt{+A+P+R} (tied with \texttt{+A}) for \mypipeline{}, \texttt{+P+R} for DIN-SQL, \texttt{+A+R} for MAC-SQL, and \texttt{+A+P+R} (tied with \texttt{+R}) for CSCSQL. The full \texttt{+A+P+R} transformation is consistently strong: it ties or beats every other configuration in roughly half of the eight pipeline-benchmark cells.
% and stays within two points of the best in the remaining cells. 
The operator-level analysis below sketches which schema-related bottleneck each operator addresses.

% \noindent\textbf{Schema abstraction.}
% Schema abstraction is most helpful when the pipeline struggles with recurring multi-table access patterns. In these cases, semantic views reduce the burden of join-path inference by exposing higher-level logical access paths directly in the schema representation. This effect is particularly visible in several BIRD-Union and Spider-Union settings where $+A$ alone is already sufficient to boost the performance, although schema abstraction is not uniformly dominant across all pipelines.

\noindent\textbf{Schema abstraction (+A).} Schema abstraction is particularly effective when complex, recurring join paths dominate the error space. For example, applying \texttt{+A} alone to \mypipeline{} (using GPT-5.4 mini on Spider-Union) improves EX from 81.7\% to 84.5\%. 
This gain is not merely a uniform improvement. JOIN-path errors account for 29.3\% of the 92 baseline failures, and +A repairs these errors at a 33.3\% rate, which is 44\% higher than its repair rate on non-JOIN failures (23.1\%), empirically confirming that abstraction specifically targets errors caused by complex join structures.

% \noindent\textbf{Schema partitioning.}
% Cluster-based organization becomes relevant in the history-aware setting, where it helps restrict the retrieval of relevant query examples and reasoning to the most relevant schema region. Strong gains from configurations such as \texttt{R+P}, \texttt{P+A}, and \texttt{R+A+P} (especially when comparing \texttt{R+A+P} with \texttt{R+A}) suggest that schema partitioning and workload-aware organization can provide benefits beyond raw access to historical SQL alone.

\noindent\textbf{Schema partitioning (+P).}
+P is most useful in the history-aware setting because it changes both the schema context and the demonstration retrieval pool. Its benefit is clearest when adding +P to an existing combination improves performance, such as \texttt{+A+R} versus \texttt{+A+P+R} or \texttt{A} versus \texttt{+A+P}, suggesting that schema partitioning and workload-aware organization can provide benefits beyond raw access to historical SQL alone.

% \noindent\textbf{Schema Renaming.}
% Rename-based transformations are often effective when the dominant difficulty lies in weak identifier semantics. This is particularly visible in many settings on both BIRD-Union and Spider-Union when comparing the configurations $+R+A+P$ and $+P+A$. These results support the intuition that clearer identifiers improve lexical alignment between user questions and schema elements.

\noindent\textbf{Schema renaming (+R).}
+R helps when the bottleneck is lexical alignment between user questions and schema elements. We analyze its effect by comparing configurations with and without \texttt{+R}, such as \texttt{+A+P} versus \texttt{+A+P+R}. The gains are not uniform, suggesting that renaming is most useful when identifier ambiguity is a dominant failure mode, but can be redundant or noisy otherwise. On Spider-Union with Gemini 2.5, \texttt{+R} alone yields a 0.8 percentage-point EX gain. Among the cases corrected by renaming, Table-set Match rises from 64.7\% to 94.1\%, while filter and predicate-related metrics remain largely unchanged. This suggests that renaming primarily helps by improving lexical schema alignment and establishing the table-level grounding needed for later stages to construct correct predicates and joins.

\subsection{Robustness Across Backbones and Workload-History Availability}
\label{sec:robustness}
\textbf{The gains of model-facing schema persist across LLM backbones and across the spectrum of historical workload availability.} Two robustness dimensions support this.

\noindent\textbf{Across backbones.}
Table~\ref{tab:bird-union-with-history-backbone} isolates the effect of LLM backbone choice for \mypipeline{}. Across both BIRD-Union and Spider-Union, transformed schemas improve performance with GPT-4.1 mini, Gemini 2.5-flash-lite, and GPT-5.4 mini. The results show that the most effective transformation depends on the backbone. For GPT-4.1 mini and Gemini 2.5-flash-lite, the largest improvements generally come from combined transformations such as \texttt{+A+P} or \texttt{+A+P+R}, suggesting that these models benefit from simultaneously improving identifier clarity, schema partitioning, and join-path abstraction. In contrast, GPT-5.4 mini obtains its strongest gains from schema abstraction alone (\texttt{+A}), suggesting that stronger models may already handle some forms of lexical ambiguity and schema partitioning, while still benefiting from higher-level logical access paths. 
% These findings reinforce our broader conclusion that schema transformations are complementary but non-monotonic: different backbones benefit from different parts of the transformation space.

\begin{figure}[t]
    \centering
    % \linewidth locks the image to the boundaries of one column
    \includegraphics[width=\linewidth]{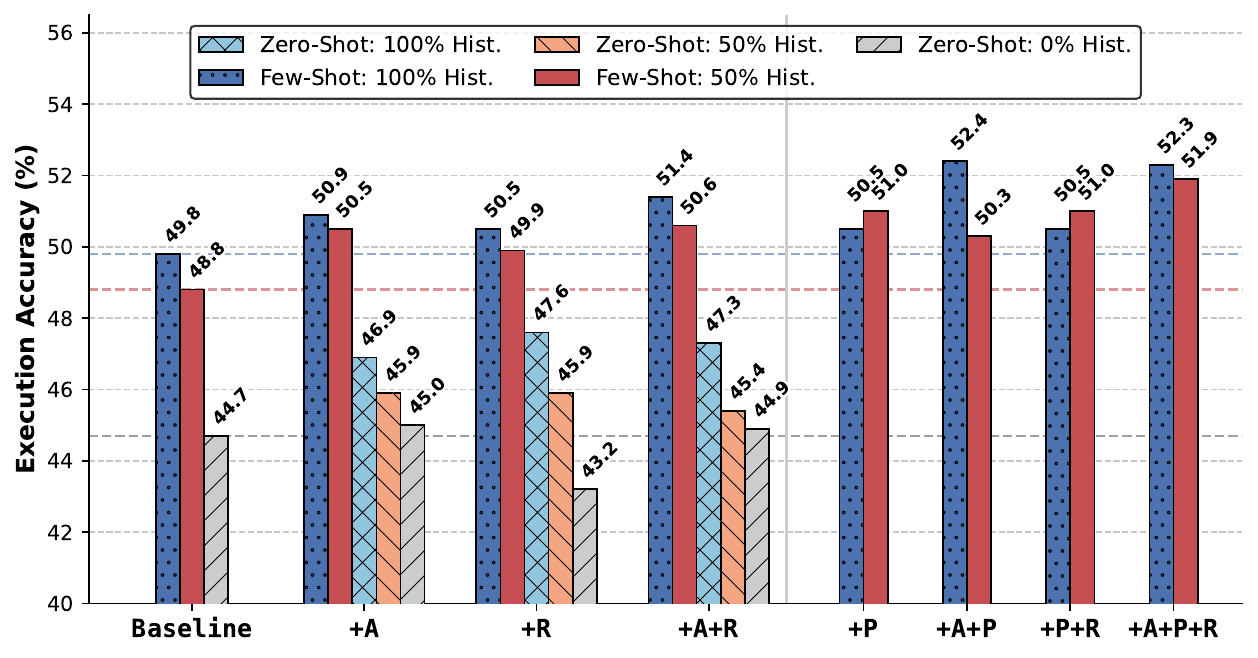}
    %\vspace{-15pt} % 1. Pulls the caption up closer to the X-axis of the chart
    % \caption{Ablation study of schema transformations on BIRD-Union. (R: Rename, A: Abstraction, P: Partitioning)}
    \caption{Execution accuracy on BIRD-Union under 0/50/100\% historical workload availability by transformation.}
    \label{fig:BIRD-union-ablation}
    %\vspace{-5pt} % 2. Pulls the NEXT figure (Figure 8) up closer to this caption
\end{figure}

\begin{figure}[t]
    \centering
    %\vspace{-5pt} % 3. Pulls this chart slightly closer to the caption of Figure 7 above it
    \includegraphics[width=\linewidth]{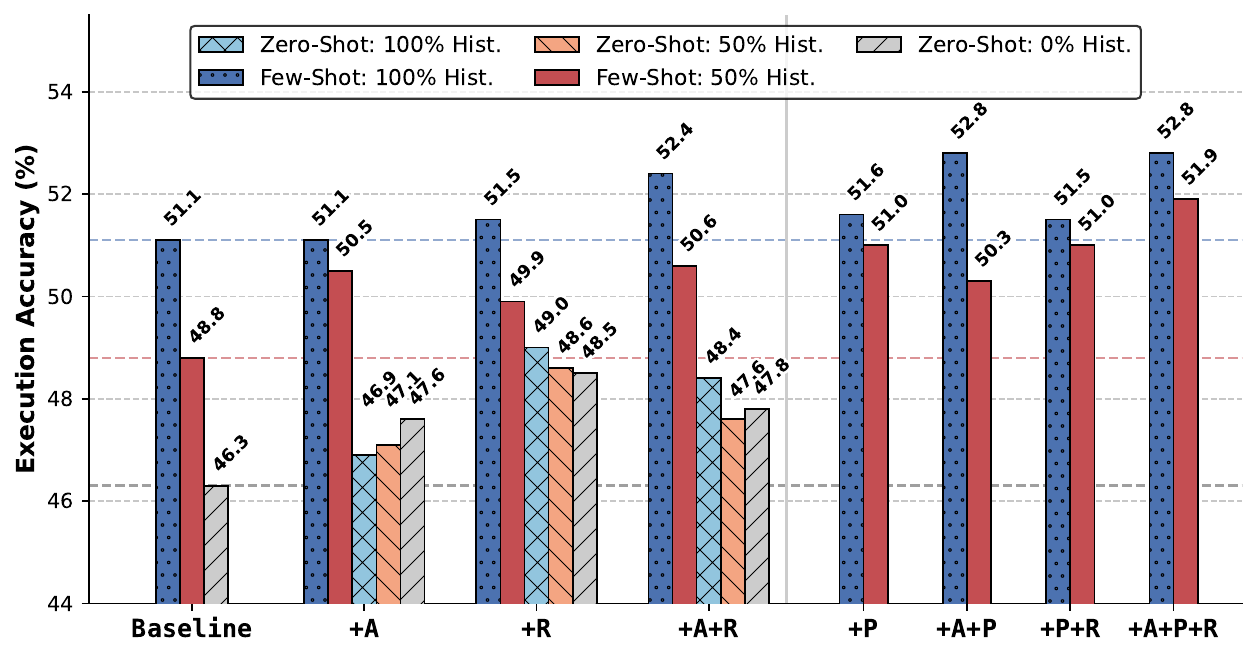}
    %\vspace{-15pt} % Pulls caption up
    % \caption{Ablation study of schema transformations (per-db) on BIRD-Union. (R: Rename, A: Abstraction, P: Partitioning)}
    \caption{Per-database execution accuracy on BIRD under 0/50/100\% historical workload availability by transformation.}
    \label{fig:BIRD-perdb-ablation}
    %\vspace{-5pt} % Pulls the NEXT figure (Figure 9) up closer
\end{figure}

\noindent\textbf{Across workload-history availability.}
Figure~\ref{fig:BIRD-union-ablation} and Figure~\ref{fig:BIRD-perdb-ablation} present how performance changes as the amount of available historical workload increases (0\%, 50\%, 100\%). We evaluate these variations in design-time history across both zero-shot (no prompt-time history) and few-shot (with prompt-time history) settings. Configurations containing \texttt{+P} are omitted from zero-shot evaluation as partitioning requires design-time history to construct partitions, and applying it without retrieving the associated prompt-time demonstrations would be impractical. The results show two consistent trends. First, schema transformation remains beneficial even with 0\% design-time history (when comparing the gray bars for optimizations like \texttt{+A} and \texttt{+A+R} with the baseline), confirming that the framework is effective without any historical workload. Second, performance generally improves as more historical workload data becomes available, indicating that workload-aware schema transformation and retrieval yield additional gains when such information is available. The conclusion is similar when we consider the BIRD-per-DB setting (the original BIRD benchmark but still no evidence or column description), where the main difference is that the gaps between the 50\% and 100\% design-time history settings are larger than on BIRD-Union (when comparing the dark blue bars with the red bars). Schema transformation is therefore not tied to having a workload history, functioning as a viable optimization even in completely history-free environments.

\subsection{Per-Domain Analysis and Boundary Conditions}
\label{sec:per-domain-analysis}

\begin{table}[ht]
\centering
\caption{Database characteristics sorted by baseline performance. $N_{1-3}$ denote the ratio of column identifier naturalness~\cite{snail} ($N_1$: most, $N_3$: least).}
\label{tab:db_stats}
\setlength{\tabcolsep}{3pt} % Tighten horizontal space
% \footnotesize % Slightly smaller font to fit comfortably
\resizebox{\columnwidth}{!}{%
\begin{tabular}{l cccccc r}
\toprule
\textbf{Database Domain} & \textbf{Tbls} & \textbf{Cols} & \textbf{$N_1$} & \textbf{$N_2$} & \textbf{$N_3$} & \textbf{FKs} & \textbf{Qrys} \\
\midrule
thrombosis\_prediction   & 3  & 64  & 18.8\% & 4.7\%  & 76.6\% & 2  & 82 \\
california\_schools      & 3  & 89  & 50.6\% & 44.9\% & 4.5\%  & 2  & 47 \\
financial                & 8  & 55  & 60.0\% & 12.7\% & 27.3\% & 8  & 46 \\
european\_football\_2    & 7  & 199 & 51.3\% & 33.7\% & 15.1\% & 30 & 58 \\
formula\_1               & 13 & 94  & 88.3\% & 7.4\%  & 4.3\%  & 19 & 86 \\
toxicology               & 4  & 11  & 72.7\% & 27.3\% & 0.0\%  & 5  & 81 \\
card\_games               & 6  & 115 & 73.9\% & 25.2\% & 0.9\%  & 4  & 93 \\
debit\_card\_specializing & 5  & 21  & 95.2\% & 4.8\%  & 0.0\%  & 2  & 34 \\
student\_club            & 8  & 48  & 100\%  & 0.0\%  & 0.0\%  & 8  & 76 \\
superhero                & 10 & 31  & 67.7\% & 32.3\% & 0.0\%  & 11 & 59 \\
codebase\_community      & 8  & 71  & 88.7\% & 11.3\% & 0.0\%  & 13 & 105 \\
\bottomrule
\end{tabular}%
}
\end{table}

\begin{figure}[t]
    \centering
    %\vspace{-5pt} % Pulls this chart slightly closer to the caption of Figure 8 above it
    \includegraphics[width=\linewidth]{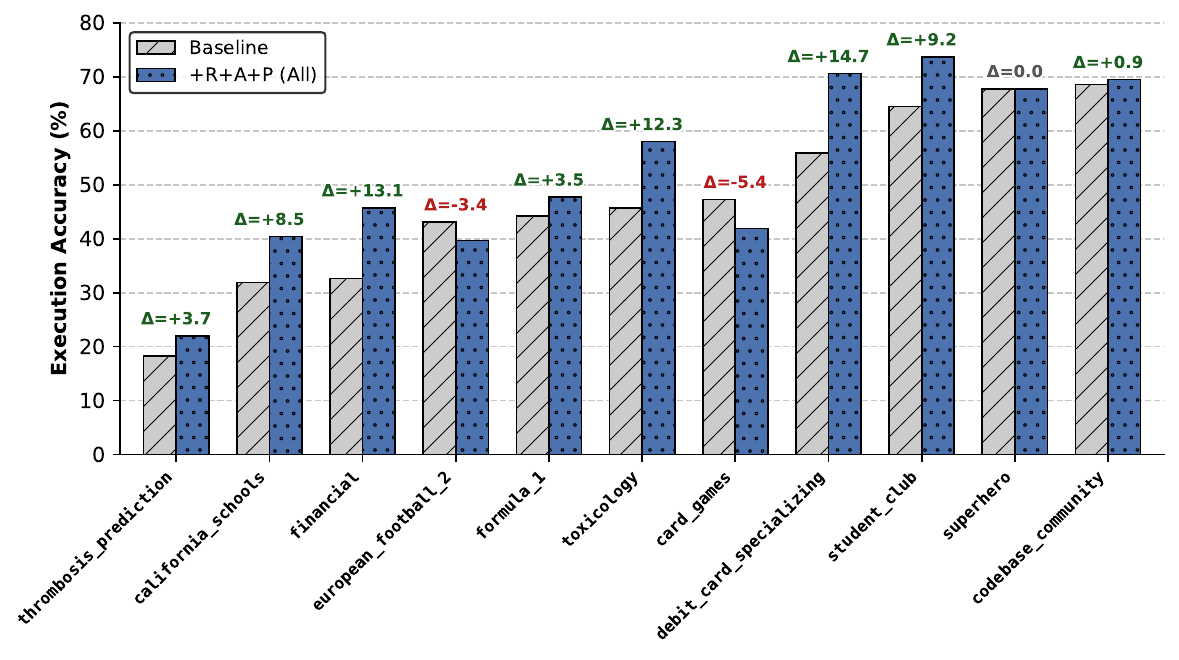}
    %\vspace{-15pt} % Pulls caption up
    \caption{Database-level breakdown of execution accuracy on BIRD-Union: baseline vs.\ the full \texttt{+A+P+R} transformation.}
    \label{fig:db_ablation_single}
    %\vspace{-5pt} % 4. Aggressively pulls the body text below Figure 9 upward
\end{figure}

Table~\ref{tab:db_stats} and Figure~\ref{fig:db_ablation_single} provide a database-level breakdown on BIRD-Union, evaluated using \mypipeline{} with Gemini 2.5-flash-lite. Overall, the transformed schema improves performance on 8 out of 11 databases, leaves one unchanged, and degrades performance on two. This shows that logical schema transformation is broadly useful, but not uniformly beneficial. The largest gains occur in domains where the dominant difficulty appears to be schema-level reasoning, such as weak identifier semantics, noisy schema context, or nontrivial relational access paths. For example, gains on domains such as \texttt{toxicology}, \texttt{student\_club}, and \texttt{formula\_1} suggest that schema abstraction and partitioning can help expose recurring access patterns and reduce irrelevant context.

The negative cases reveal the boundary conditions of the approach. The two queries below illustrate the failure mode: the renamed schema's identifiers (e.g., \texttt{Bank\_Clients}, \texttt{Energy\_Stations}) are plausible aliases but point the model to the wrong table for the question, leading to a confidently wrong join.

\begin{tcolorbox}[SQLbox, breakable, before skip=4pt, after skip=4pt]
\begin{lstlisting}[style=psqlcolor,mathescape,deletendkeywords={YEAR, DATE}]
Example A (financial). In the branch where the second-highest number of crimes were committed in 1995 occurred, how many male clients are there?  
$Q_{corr}=$ SELECT COUNT(T1.client_id) FROM client AS T1  WHERE 
T1.gender = 'M' AND T1.district_id = (SELECT T2.district_id 
FROM account AS T2 WHERE STRFTIME('%Y', T2.date) = '1995' 
GROUP BY T2.district_id ORDER BY COUNT(*) DESC 
LIMIT 1 OFFSET 1);

$Q_{wrong}=$ SELECT COUNT(T1.client_id) FROM `Bank_Clients` 
AS T1 INNER JOIN `Bank_Districts` AS T2  
ON T1.district_id = T2.district_id WHERE T2.district_id = 
(SELECT district_id FROM `Bank_Districts` ORDER BY
`crime_count_95` DESC LIMIT 1 OFFSET 1) AND T1.gender = 'M';
\end{lstlisting}

\begin{lstlisting}[style=psqlcolor,mathescape,deletendkeywords={YEAR, DATE}]
Example B (debit_card_specializing).  What is the nationality of the customer who spent 548.4 in 2012/8/24?  
$Q_{corr}=$ SELECT T2.Segment FROM transactions_1k AS T1 
INNER JOIN customers AS T2 ON T1.CustomerID = T2.CustomerID 
WHERE T1.Date = '2012-08-24' AND T1.Price = 548.4;

$Q_{wrong}=$ SELECT T2.country_code FROM Energy_Sales AS T1 
INNER JOIN Energy_Stations AS T2 ON T1.station_id = 
T2.station_id WHERE T1.trans_date = '2012-08-24' 
AND T1.total_price = 548.4;
\end{lstlisting}
\end{tcolorbox}

When errors are driven less by schema structure and more by domain-specific value grounding, subtle predicate interpretation, or fine-grained column selection, schema transformation may provide limited benefit or even hurt. For example, on \texttt{thrombosis\_}\allowbreak\texttt{prediction}, performance remains unchanged despite a high proportion of complex queries and less natural schema identifiers, suggesting that the remaining errors may be dominated by medical terminology, value grounding, or predicate semantics rather than by schema transformation alone. In such cases, abstraction may introduce unhelpful alternatives, partitioning may prune useful context, or renaming may remove lexical cues present in the original schema. 
The largest degradation occurs on \texttt{card\_games} ($\Delta=-5.4$), a wide schema with many semantically dense attributes but relatively low average join complexity. These results refine the operator-level conclusion: Text-to-SQL-friendly logical design is useful but non-monotonic, and no single transformation is uniformly optimal across schemas.

\begin{figure}[t]
    \centering
    % \linewidth locks the image to the boundaries of one column
    \includegraphics[width=\linewidth]{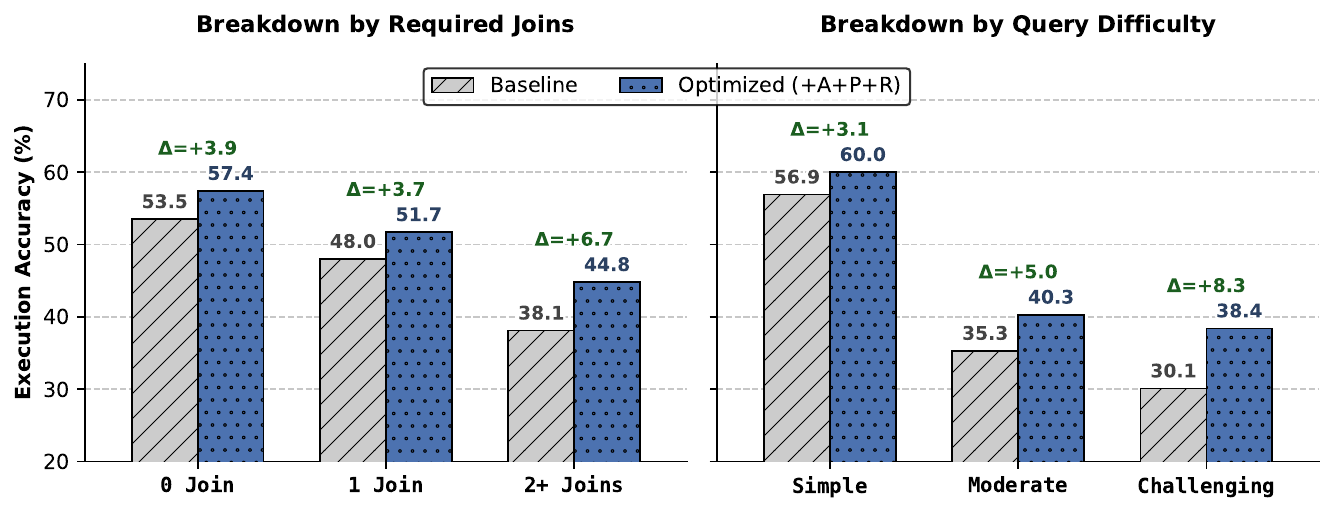}
    %\vspace{-20pt} % Pulls caption up
    \caption{Performance breakdown by joins and difficulty}
    \label{fig:join-difficulty-breakdown}
    %\vspace{-5pt} % Pulls caption up
\end{figure}

\smallskip
\noindent\textbf{Performance breakdown by query complexity.}
Figure ~\ref{fig:join-difficulty-breakdown} reports performance breakdowns on BIRD-Union by required joins and by the query difficulty labels provided by the BIRD benchmark. This diagnostic analysis decomposes the same experimental results evaluated in Section~\ref{sec:per-domain-analysis}. The goal is to understand where the complete schema-optimization pipeline improves over the baseline.
The results show that schema transformation improves both single-table and join-requiring queries, with slightly larger gains for join-requiring queries. The effect is more pronounced in the difficulty-based breakdown: gains increase from simple to moderate, then to challenging queries. This suggests that logical schema transformation is especially helpful when the model faces a greater schema-induced reasoning burden.

\subsection{Error Analysis}

\begin{figure}[t]
    \centering
    % \linewidth locks the image to the boundaries of one column
    \includegraphics[width=\linewidth]{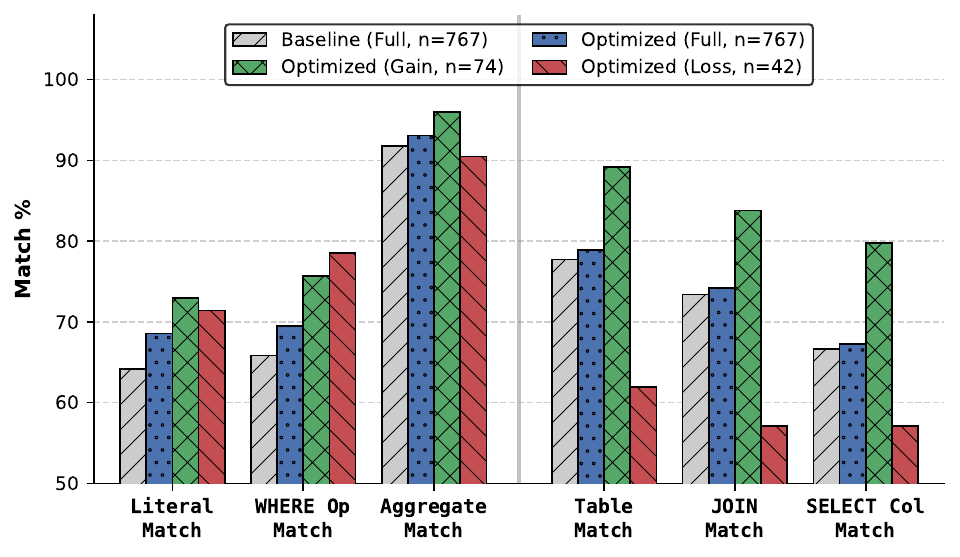}
    %\vspace{-20pt}
    \caption{Error Analysis on BIRD-Union}
    \label{fig:error-analysis}
    %\vspace{-5pt}
\end{figure}

Figure~\ref{fig:error-analysis} utilizes the same result evaluated in Section~\ref{sec:per-domain-analysis} to analyze the +4.2\% execution accuracy gain across six SQL component-match metrics. It contrasts the baseline and optimized pipelines over the full evaluation set ($n{=}767$), as well as the specific ``gain'' ($n{=}74$) and ``loss'' ($n{=}42$) query subsets, where the optimized pipeline's outcome differs from the baseline. Overall, the optimized pipeline improves full-set performance across all metrics (by 0.8 to 4.2\%). However, the impact is highly asymmetric across SQL abstraction levels. On fine-grained components (Literal, WHERE Operators, Aggregate Functions), the optimized pipeline achieves its most significant full-set gains, while the gain and loss subsets diverge by at most 5\%. This demonstrates that the optimized schema consistently enhances filter-level reasoning via better semantic connections, regardless of ultimate query success. Conversely, on coarse-grained structural components (Table, JOIN Conditions, SELECT Columns), the gain and loss subsets diverge drastically (by 22 to 27\%), with the loss subset plummeting well below baseline performance. This clearly showcases the failure mode: these 42 losses stem almost entirely from aggressive schema partitioning inadvertently pruning a required table or from the retrieval of an incorrect view. Ultimately, the net improvement of 32 queries is achieved because fine-grained semantic errors dominate the baseline's failure space, and the broad consistency gained from resolving these semantic gaps significantly outweighs the framework's penalties.

\section{Conclusion and Future Work}

In this paper, we introduced Text-to-SQL-friendly logical database design, a novel optimization framework that shifts the reasoning burden from the language model's prompt to the underlying schema representation. By achieving this objective through three transformations---Schema Abstraction (\texttt{+A}), Schema Partitioning (\texttt{+P}), and Schema Renaming (\texttt{+R})---we demonstrated that optimizing the model-facing schema yields pipeline-agnostic performance gains. Our framework seamlessly integrates with existing Text-to-SQL systems, delivering consistent accuracy improvements across multiple LLM backbones in both zero-shot and few-shot settings. Ultimately, as the primary consumer of the schema shifts from human developers to language models, Text-to-SQL-friendly logical design will become just as fundamental an objective as classical normalization and data integrity.

Moving forward, there are several promising avenues for extending this work. First, future frameworks could focus on dynamic, cost-aware view materialization for massive-scale databases. Second, subsequent research could explore context-aware schema renaming tailored to the user's specific phrasing. Finally, developing adaptive routing mechanisms will be critical to mitigate domain-specific regressions.

\bibliographystyle{ACM-Reference-Format}
\bibliography{references}

@inproceedings{yu2018spider,
  author    = {Yu, Tao and Zhang, Rui and Yang, Kai and Yasunaga, Michihiro  and
               Wang, Dongxu and Li, Zifan and Ma, James and Li, Irene and
               Yao, Qingning  and Roman, Shanelle and Zhang, Zilin and
               Radev, Dragomir},
  title     = {Spider: A Large-Scale Human-Labeled Dataset for Complex and Cross-Domain Semantic Parsing and Text-to-SQL Task},
  booktitle = {Proceedings of the 2018 Conference on Empirical Methods in Natural Language Processing},
  year      = {2018}
}

@inproceedings{li2023bird,
  author    = {Li, Jinyang and Hui, Binyuan and Qu, Ge and Yang, Jiaxi and
               Li, Binhua and Li, Bowen and Wang, Bailin and Qin, Bowen and
               Cao, Rongyu and Geng, Ruiying and Huo, Nan and Zhou, Xuanhe and
               Ma, Chenhao and Li, Guoliang and Chang, Kevin C. C. and
               Huang, Fei and Cheng, Reynold and Li, Yongbin },
  title     = {Can LLM Already Serve as A Database Interface? A Big Bench for Large-Scale Database Grounded Text-to-SQLs},
  booktitle = {Advances in Neural Information Processing Systems},
  volume    = {36},
  year      = {2023}
}

@inproceedings{chaudhuri1997autoadmin,
  author    = {Chaudhuri, Surajit and Narasayya, Vivek},
  title     = {An Efficient, Cost-Driven Index Selection Tool for Microsoft SQL Server},
  booktitle = {Proceedings of the 23rd International Conference on Very Large Data Bases (VLDB)},
  year      = {1997}
}

@article{snail,
  title      = {{SNAILS: Schema Naming Assessments for Improved LLM-Based SQL Inference}},
  author     = {Luoma, Kyle and Kumar, Arun},
  journal    = {Proceedings of the ACM on Management of Data (PACMMOD)},
  volume     = {3},
  number     = {1},
  articleno  = {77},
  pages      = {1--26},
  year       = {2025},
  month      = {February},
  publisher  = {Association for Computing Machinery},
  doi        = {10.1145/3709727}
}

@inproceedings{zhao2022schemaexpansion,
  title     = {Bridging the Generalization Gap in Text-to-{SQL} Parsing with Schema Expansion},
  author    = {Zhao, Chen and Su, Yu and Pauls, Adam and Platanios, Emmanouil Antonios},
  booktitle = {Proceedings of the 60th Annual Meeting of the Association for Computational Linguistics (ACL)},
  pages     = {5568--5578},
  year      = {2022},
  doi       = {10.18653/v1/2022.acl-long.381},
  url       = {https://aclanthology.org/2022.acl-long.381/}
}

@misc{bmc22022sqlcreatecontext,
  title        = {{SQL Create Context}: A Dataset with Attribute-Annotated {DDL} for Text-to-SQL},
  author       = {{b-mc2} Contributors},
  howpublished = {Hugging Face dataset},
  year         = {2022},
  url          = {https://huggingface.co/datasets/b-mc2/sql-create-context}
}

@inproceedings{talaei2025chess,
  title     = {{CHESS}: Contextual Harnessing for Efficient SQL Synthesis},
  author    = {Talaei, Shayan and Pourreza, Mohammadreza and Chang, Yu-Chen and Mirhoseini, Azalia and Saberi, Amin},
  booktitle = {ICML 2025 Workshop on Multi-Agent Systems in the Era of Foundation Models},
  year      = {2025}
}

@inproceedings{furst2025evaluating,
  author    = {Jonathan F\"{u}rst and Catherine Kosten and Farhad Nooralahzadeh and Yi Zhang and Kurt Stockinger},
  title     = {Evaluating the Data Model Robustness of Text-to-SQL Systems Based on Real User Queries},
  booktitle = {Proceedings of the 28th International Conference on Extending Database Technology (EDBT)},
  year      = {2025}
}

@inproceedings{wang2025mac,
  title     = {MAC-SQL: A Multi-Agent Collaborative Framework for Text-to-SQL},
  author    = {Wang, Bing and Ren, Changyu and Yang, Jian and Liang, Xinnian and Bai, Jiaqi and Chai, Linzheng and Yan, Zhao and Zhang, Qian-Wen and Yin, Di and Sun, Xing and others},
  booktitle = {Proceedings of the 31st International Conference on Computational Linguistics},
  pages     = {540--557},
  year      = {2025}
}

@inproceedings{alphasql2025,
  title     = {{Alpha-SQL}: Zero-Shot Text-to-SQL using Monte Carlo Tree Search},
  author    = {Li, Boyan and Zhang, Jiayi and Fan, Ju and Xu, Yanwei and Chen, Chong and Nan, Tang and Luo, Yuyu},
  booktitle = {Proceedings of the 42nd International Conference on Machine Learning (ICML)},
  year      = {2025}
}

@article{evoschema2025,
  title     = {{EvoSchema}: Towards Text-to-{SQL} Robustness Against Schema Evolution},
  author    = {Zhang, Tianshu and Qian, Kun and Sahai, Siddartha and Tian, Yuan and Garg, Shaddy and Sun, Huan and Li, Yunyao},
  journal   = {Proceedings of the VLDB Endowment},
  volume    = {18},
  number    = {10},
  pages     = {3655--3668},
  year      = {2025},
  publisher = {VLDB Endowment}
}

@inproceedings{reforce2025,
  title     = {{ReFoRCE}: A Text-to-SQL Agent with Self-Refinement, Format Restriction, and Column Exploration},
  author    = {Deng, Minghang and Ramachandran, Ashwin and Xu, Canwen and Hu, Lanxiang and Yao, Zhewei and Datta, Anupam and Zhang, Hao},
  booktitle = {ICLR 2025 Workshop on VerifAI},
  year      = {2025}
}

@inproceedings{deepeyesql2026,
  title   = {{DeepEye-SQL}: A Software-Engineering-Inspired Text-to-SQL Framework},
  author  = {Li, Boyan and Chen, Chong and Xue, Zhujun and Mei, Yinan and Luo, Yuyu},
  journal = {Proceedings of the ACM on Management of Data},
  year    = {2026}
}

@article{eben2025rasl,
  author     = {Jeffrey Eben and Aitzaz Ahmad and Stephen Lau},
  title      = {{RASL:} Retrieval Augmented Schema Linking for Massive Database Text-to-SQL},
  journal    = {CoRR},
  volume     = {abs/2507.23104},
  year       = {2025},
  url        = {https://doi.org/10.48550/arXiv.2507.23104},
  doi        = {10.48550/ARXIV.2507.23104},
  eprinttype = {arXiv},
  eprint     = {2507.23104}
}

@inproceedings{xu2025tssql,
  title     = {{TS}-{SQL}: Test-driven Self-refinement for Text-to-{SQL}},
  author    = {Xu, Wenbo and Zhu, Haifeng and Yan, Liang and Liu, Chuanyi and Han, Peiyi and Duan, Shaoming and Pan, Jeff Z.},
  booktitle = {Findings of the Association for Computational Linguistics: EMNLP 2025},
  pages     = {2864--2889},
  year      = {2025},
  publisher = {Association for Computational Linguistics}
}

@article{xie2025opensearchsql,
  title     = {OpenSearch-SQL: Enhancing Text-to-SQL with Dynamic Few-shot and Consistency Alignment},
  author    = {Xie, Xiangjin and Xu, Guangwei and Zhao, Lingyan and Guo, Ruijie},
  journal   = {Proceedings of the ACM on Management of Data},
  volume    = {3},
  number    = {3},
  articleno = {194},
  pages     = {1--24},
  year      = {2025},
  doi       = {10.1145/3725331},
  url       = {https://dl.acm.org/doi/10.1145/3725331}
}

@article{cao2024rsl,
  title   = {{RSL-SQL}: Robust Schema Linking in Text-to-SQL Generation},
  author  = {Cao, Zhenbiao and Zheng, Yuanlei and Fan, Zhihao and Zhang, Xiaojin and Chen, Wei and Bai, Xiang},
  journal = {arXiv preprint arXiv:2411.00073},
  year    = {2024},
  doi     = {10.48550/arXiv.2411.00073},
  url     = {https://arxiv.org/abs/2411.00073}
}

@article{li2025omnisql,
  title   = {{OmniSQL}: Synthesizing High-Quality Text-to-SQL Data at Scale},
  author  = {Li, Haoyang and Wu, Shang and Zhang, Xiaokang and Huang, Xinmei and Zhang, Jing and Jiang, Fuxin and Wang, Shuai and Zhang, Tieying and Chen, Jianjun and Shi, Rui and Chen, Hong and Li, Cuiping},
  journal = {Proceedings of the VLDB Endowment},
  volume  = {18},
  number  = {11},
  pages   = {4695--4709},
  year    = {2025},
  doi     = {10.14778/3749646.3749723},
  url     = {https://dl.acm.org/doi/10.14778/3749646.3749723}
}

@inproceedings{pourreza2023dinsql,
  title     = {{DIN-SQL}: Decomposed In-Context Learning of Text-to-{SQL} with Self-Correction},
  author    = {Pourreza, Mohammadreza and Rafiei, Davood},
  booktitle = {Advances in Neural Information Processing Systems (NeurIPS)},
  volume    = {36},
  pages     = {28448--28472},
  year      = {2023}
}

@article{gao2024dailsql,
  title     = {Text-to-{SQL} Empowered by Large Language Models: A Benchmark Evaluation},
  author    = {Gao, Dawei and Wang, Haibin and Li, Yaliang and Sun, Xiuyu and Qian, Yichen and Ding, Bolin and Zhou, Jingren},
  journal   = {Proceedings of the VLDB Endowment},
  volume    = {17},
  number    = {5},
  pages     = {1132--1145},
  year      = {2024},
  publisher = {VLDB Endowment}
}

@inproceedings{CSCSQL,
  author    = {Sheng, Lei and Shuai, Xu Shuai},
  title     = {{CSC-SQL}: Corrective Self-Consistency in Text-to-SQL via Reinforcement Learning},
  booktitle = {Findings of the Association for Computational Linguistics: IJCNLP-AACL 2025},
  pages     = {1473--1496},
  year      = {2025},
  publisher = {Association for Computational Linguistics},
  doi       = {10.18653/v1/2025.findings-ijcnlp.91},
  url       = {https://aclanthology.org/2025.findings-ijcnlp.91/}
}

@book{garcia-molina-dbsystems,
  author    = {Garcia-Molina, Hector and Ullman, Jeffrey D. and Widom, Jennifer},
  title     = {Database Systems: The Complete Book},
  edition   = {2},
  publisher = {Pearson},
  year      = {2008}
}

@article{Halevy2001,
  author  = {Halevy, Alon Y.},
  title   = {Answering queries using views: A survey},
  journal = {The VLDB Journal},
  year    = {2001},
  volume  = {10},
  pages   = {270--294},
  doi     = {10.1007/s007780100054}
}

@article{chaudhuri2007selftuning,
  author  = {Surajit Chaudhuri and Gerhard Weikum},
  title   = {Foundations of Automated Database Tuning},
  journal = {Proceedings of the VLDB Endowment},
  volume  = {1},
  number  = {1},
  pages   = {3--14},
  year    = {2008}
}

@inproceedings{kotidis1999dynamat,
  author    = {Kotidis, Yannis and Roussopoulos, Nick},
  title     = {{DynaMat}: A Dynamic View Management System for Data Warehouses},
  booktitle = {Proceedings of the 1999 ACM SIGMOD International Conference on Management of Data},
  pages     = {371--382},
  year      = {1999}
}

@inproceedings{harinarayan1996implementing,
  author    = {Venky Harinarayan and Anand Rajaraman and Jeffrey D. Ullman},
  title     = {Implementing Data Cubes Efficiently},
  booktitle = {Proceedings of the {ACM} {SIGMOD} International Conference on Management of Data},
  pages     = {205--216},
  publisher = {{ACM} Press},
  year      = {1996},
  url       = {https://doi.org/10.1145/233269.233333},
  doi       = {10.1145/233269.233333}
}

@inproceedings{agrawal2000automated,
  author    = {Sanjay Agrawal and Surajit Chaudhuri and Vivek R. Narasayya},
  title     = {Automated Selection of Materialized Views and Indexes in {SQL} Databases},
  booktitle = {Proceedings of the 26th International Conference on Very Large Data Bases ({VLDB})},
  pages     = {496--505},
  publisher = {VLDB Endowment},
  year      = {2000},
  url       = {http://www.vldb.org/conf/2000/P496.pdf}
}

@inproceedings{zilio2004db2,
  author    = {Daniel C. Zilio and Jun Rao and Sam Lightstone and Guy M. Lohman and Adam J. Storm and Christian Garcia{-}Arellano and Scott Fadden},
  title     = {{DB2} Design Advisor: Integrated Automatic Physical Database Design},
  booktitle = {Proceedings of the 30th International Conference on Very Large Data Bases ({VLDB})},
  pages     = {1087--1097},
  publisher = {VLDB Endowment},
  year      = {2004},
  url       = {http://www.vldb.org/conf/2004/IND4P1.PDF},
  doi       = {10.1016/B978-012088469-8.50095-4}
}

@inproceedings{agrawal2004autopart,
  author    = {Sanjay Agrawal and Vivek R. Narasayya and Beverly Yang},
  title     = {Integrating Vertical and Horizontal Partitioning Into Automated Physical Database Design},
  booktitle = {Proceedings of the {ACM} {SIGMOD} International Conference on Management of Data},
  pages     = {359--370},
  publisher = {{ACM}},
  year      = {2004},
  url       = {https://doi.org/10.1145/1007568.1007609},
  doi       = {10.1145/1007568.1007609}
}

@article{curino2010schism,
  author  = {Carlo Curino and Yang Zhang and Evan P. C. Jones and Samuel Madden},
  title   = {Schism: a Workload-Driven Approach to Database Replication and Partitioning},
  journal = {Proceedings of the VLDB Endowment},
  volume  = {3},
  number  = {1},
  pages   = {48--57},
  year    = {2010},
  url     = {https://vldb.org/pvldb/vol3/R04.pdf},
  doi     = {10.14778/1920841.1920853}
}

@inproceedings{pavlo2017selfdriving,
  author    = {Andrew Pavlo and Gustavo Angulo and Joy Arulraj and Haibin Lin and Jiexi Lin and Lin Ma and Prashanth Menon and Todd C. Mowry and Matthew Perron and Ian Quah and Siddharth Santurkar and Anthony Tomasic and Skye Toor and Dana Van Aken and Ziqi Wang and Yingjun Wu and Ran Xian and Tieying Zhang},
  title     = {Self-Driving Database Management Systems},
  booktitle = {Proceedings of the 8th Biennial Conference on Innovative Data Systems Research ({CIDR})},
  publisher = {www.cidrdb.org},
  year      = {2017},
  url       = {http://cidrdb.org/cidr2017/papers/p42-pavlo-cidr17.pdf}
}

\newpage

\appendix

\end{document}